\documentclass[12pt,a4paper]{article}
\usepackage{graphicx,xcolor,tcolorbox}
\usepackage[T1]{fontenc}
\usepackage[utf8]{inputenc}
\usepackage{dutchcal}
\usepackage{textcomp}
\usepackage[sc,osf]{mathpazo}
\usepackage{bbm}
\usepackage{a4wide}  
\usepackage{latexsym,amsthm,amsfonts,amsmath,mathrsfs,amssymb}
\usepackage{accents}
\usepackage[nosort]{cite}
\usepackage{booktabs} 
\usepackage[unicode,implicit,colorlinks=true]{hyperref}  

\definecolor{carminered}{rgb}{1.0, 0.0, 0.22}
\hypersetup{%
	pdftitle    = {On the quantum statistical relation and the covariant
          phase space Hamiltonian}
	pdfkeywords = {Euclidean Path Integral, Euclidean Quantum gravity,
          Covariant Phase Space, Wald Formalism, Hamiltonian, Komar},
	pdfauthor   = {Jos\'e Luis V.~Cerdeira and Tom\'as Ort\'{\i}n},
	plainpages  = true,
	colorlinks  = true,
	linkcolor= purple,
	citecolor   = cyan,
	filecolor= teal,
	urlcolor    = purple
}
\newcommand{\hepth}[1]{{\tt
		\href{http://www.arXiv.org/abs/hep-th/#1}{hep-th/#1}}}
\newcommand{\grqc}[1]{{\tt
		\href{http://www.arXiv.org/abs/gr-qc/#1}{gr-qc/#1}}}
\newcommand{\phys}[1]{{\tt
		\href{http://www.arXiv.org/abs/physics/#1}{physics/#1}}}

\newcommand{\arxiv}[1]{{\tt arXiv:\href{http://www.arXiv.org/abs/#1}{#1}}}
\newcommand{\doi}[1]{{\tt DOI:\href{http://dx.doi.org/#1}{#1}}}

\allowdisplaybreaks

\makeatletter
\@addtoreset{equation}{section}
\makeatother

\begin{document}
	
\begin{flushright}
  \small
  IFT-UAM/CSIC-26-133\\
  October 9\textsuperscript{th} 2026\\
  \normalsize
\end{flushright}

\vspace{1cm}
	
\begin{center}
		
{\Large {\bf On the quantum statistical relation\\[.3cm]
                   and\\[.3cm]
the covariant phase space Hamiltonian}}
		
\vspace{1cm}
	
\renewcommand{\thefootnote}{\alph{footnote}}
		
Jos\'e Luis V.~Cerdeira\footnote{Email: {\tt jose.verez-fraguela [at]
				estudiante.uam.es}}
and Tom\'as Ort\'{\i}n\footnote{Email: {\tt Tomas.Ortin [at] csic.es}}
		
\vspace{1cm}

\setcounter{footnote}{0}
\renewcommand{\thefootnote}{\arabic{footnote}}
\vspace{.5cm}

{\it\small Instituto de F\'{\i}sica Te\'orica UAM/CSIC\\
C/ Nicol\'as Cabrera, 13--15, C.U.~Cantoblanco, E-28049 Madrid, Spain}

\vspace{1cm}

{\bf Abstract}
\end{center}
\begin{quotation} {\small We show how the covariant phase space Hamiltonian
    can be constructed in theories in which some of the fields have gauge
    freedoms. The addition to the action of boundary terms that guarantee the
    well-posedness of a variational principle and its exact gauge invariance
    for the prescribed boundary conditions turn out to be essential in this
    construction. The same ingredients are essential in the use of the action
    in the framework of Gibbons and Hawking's Euclidean approach to Quantum
    Gravity and we show how the on-shell action constructed with a Lagrangian
    with those properties can be expressed in terms of the covariant phase
    space Hamiltonian. We compute the Hamiltonian of several theories, one of
    which has Chern--Simons terms, using our prescription and show that, in
    black-hole backgrounds, the Hamiltonian gives directly the quantum
    statistical relation including terms related to non-gravitational charges
    and chemical potentials, as expected.  }
\end{quotation}

\newpage
\pagestyle{plain}
	
\tableofcontents
	
\newpage
	
\section{Introduction}

The Covariant Phase Space (CPS) formalism pioneered by Wald and collaborators
\cite{Lee:1990nz,Wald:1993nt,Iyer:1995kg}\footnote{See also
  Refs.~\cite{Barnich:2001jy,Adami:2017phg,Compere:2018aar,Gieres:2021ekc}.}
is ideally suited to define all kinds of conserved charges in field theories
and their mutual relations. The derivation of the first law of black hole
mechanics relating the variations of the black-hole conserved charges and the
Wald entropy \cite{Wald:1993nt,Iyer:1995kg}\footnote{The work terms of the
  first law associated to the electric charge, for instance, were absent in
  these works, but they can also be determined in this formalism by treating
  properly gauge symmetries
  \cite{Prabhu:2015vua,Elgood:2020svt,Elgood:2020mdx,Elgood:2020nls}.} is a
prime example of the power of this formalism.

In theories of gravity, the charge associated to any diffeomorphism (not just
those related to time evolution) is generically known as the (CPS) Hamiltonian
$H[\xi]$. It dictates the evolution of all the fields of a theory under the
transformation generated a vector field $\xi$ and its determination is clearly
one of our main interests. However, the Hamiltonian is implicitly defined by
an equation involving its variation (its ``$\delta$-derivative'') that, as it
has been argued \cite{Iyer:1995kg,Compere:2018aar}, could not be generically
solved for $H[\xi]$ (could not be ``$\delta$-integrated'').

This problem was solved by Harlow and Wu in Ref.~\cite{Harlow:2019yfa} for
gravity without matter by taking into account the boundary terms which are
needed to have an action with well-posed variational problem, \textit{i.e.}~an
action extremized by the solutions to the classical equations of motion for
certain boundary conditions. This result is a bit surprising since,
classically, the action is just an extremely convenient way to encode
equations of motion that are well behaved under certain symmetry groups. Its
value is irrelevant and, as a matter of fact, one may just work with the
unintegrated Lagrangian. Quantum mechanically, though, the value of the action
is a most important quantity that should be extremized for classical
trajectories. Harlow and Wu's result, then, establishes a connection between
the $\delta$-integrability of the Hamiltonian and the adequacy of the action
to be used in Feynman's path integral formulation of Quantum Field Theory and,
since we are dealing with gravity, in Gibbons and Hawking's Euclidean approach
to Quantum Gravity \cite{Gibbons:1976ue,Hawking:1978jz}.

One of the main goals of this paper is to extend Harlow and Wu's results to
theories of gravity coupled to matter. All matter fields (except for true
scalars\footnote{The Higgs field is not a true scalar. It is a section of a
  bundle. In more pedestrian language, it has ``gauge freedoms'' because it
  transforms as an SU$(2)$ doublet.}) have gauge freedoms and, therefore, they
should be treated properly, as in
Refs.~\cite{Prabhu:2015vua,Elgood:2020svt,Elgood:2020mdx,Elgood:2020nls}. Classically,
the action may only be invariant under the gauge transformations of the fields
up to a total derivative but, quantum mechanically, it must be exactly gauge
invariant. As we are going to see, this classically unnecessary requirement
will allow us to $\delta$-integrate the equation defining the Hamiltonian. We
will be able to provide a closed expression for it for a generic
diff-invariant field theory and we will work out explicitly several examples
with Dirichlet boundary conditions, including one with Chern--Simons terms.

The second goal of this paper is to show that the actions with well-posed
variational principles for which we managed to $\delta$-integrate the
Hamiltonian, when rotated to Euclidean signature, can be expressed in terms of
the CPS Hamiltonian, \textit{i.e.}~as integrals over $(d-2)$-dimensional
boundaries. This extends previous results for pure gravity
\cite{Garfinkle:2000ms} to generic diff-invariant field theories and points
again to the mysterious need to deal with actions with well-posed variational
problems in order to have a closed expression for the Hamiltonian. At the same
time, it allows us to compute explicitly the Euclidean action as a boundary
integral. In this direction, we show that the result is always related to the
so-called \textit{quantum statistical relation} when Dirichlet boundary
conditions are used.
	
The paper is organized as follows: in Section~\ref{sec-Theory} we review the
basics of classical field theories (Lagrangian and equations of motion) and
the construction of conserved charges associated to the symmetries of the
Lagrangian (Noether theorems). We further study the problem of defining a well
posed variational problem in which the solutions of the equations of motion
make the action stationary. In Section~\ref{sec-Hamiltonian} we review the
construction of the covariant phase space associated to a classical field
theory and the construction of the generator of the infinitesimal diffeomorphisms
of the theory in covariant phase space (the Hamiltonian) in theories that have
gauge symmetries. This is one of our main results. In
Section~\ref{sec-EuclideanPathIntegral} we review the construction of the
Euclidean path integral and show how to express the on-shell Euclidean action
in terms of the Hamiltonian constructed in the previous section.  Finally, in
Section~\ref{sec-QSR} we show that one can recover the quantum statistical
relation using our previous results. Section~\ref{sec-Conclusions} contains
our conclusions and further directions.

\section{Theory, symmetries and charges}
\label{sec-Theory}

For our purposes here, a classical field theory in $d$ dimensions is
determined by a Lagrangian volume form\footnote{We use the conventions of
  Ref.~\cite{Ortin:2015hya} and differential-form language throughout. Most,
  but not all, of the material in this section is discussed in more depth and
  detail in, for
  instance, Refs.~\cite{Barnich:2001jy,Compere:2018aar,Adami:2017phg,Gieres:2021ekc}. Our
  main goal here is to introduce the basic concepts and the notation we are
  going to use.}  $\mathbf{L}$ depending locally on some fields which we
collectively denote by $\varphi$ and finitely many of their derivatives.  In
general the fields $\varphi$ will have one or several gauge freedoms and one
of our goals in this paper is to take into account this fact in all our
constructions.

Classically, the Lagrangian is just a very convenient way of encoding the
equations of motion, which can be derived by computing its variation
$\delta \mathbf{L}$ with respect to arbitrary infinitesimal perturbations of
the field$\varphi\mapsto \varphi+\delta \varphi$ and integrating by parts as
many times as necessary to obtain terms linear in $\delta \varphi$ and a total
derivative. The result is the \textit{first variational formula}

\begin{equation}
\label{eq:FirstVariationalFormula}
  \delta \mathbf{L}
  =
  \mathbf{E}_{\varphi}\wedge \delta \varphi
  +d\mathbf{\Theta}(\varphi,\delta \varphi)\,,
\end{equation}

By definition, the classical equations of motion for each field $\varphi$

\begin{equation}
\mathbf{E}_{\varphi} = 0\,,  
\end{equation}

\noindent
while $\mathbf{\Theta}(\varphi,\delta \varphi)$ is known as the
\textit{presymplectic potential} of the theory. Often, we will refer to the
left-hand sides of the equations of motion, $\mathbf{E}_{\varphi}$ as to
equations of motion.

This definition immediately implies that the addition of total derivatives to
a Lagrangian does not change the equations of motion. 

Although the above definition of the equations of motion can be motivated by a
variational (minimal or stationary action) principle, at the classical level
it is sufficient to define and study the dynamics of classical field
theories. The value of the action for a field configuration is not relevant
at the purely classical level. At the quantum-mechanical level and in the
context of Feynman's path-integral approach, it is an essential quantity whose
value has to be extremized by the solutions to the classical equations of
motion. We will discuss the construction of actions with this property from a
given Lagrangian in Section~\ref{sec-VariationalProblem}. Before that, we
discuss the notion of symmetry and the construction of charges in this
classical context.

\subsection{Symmetries and charges}
\label{sec-symmetries}

If under some transformations of the fields with parameters $\Lambda$,
$\delta_{\Lambda}\varphi$

\begin{equation}
  \label{eq:GaugeTransformationLagrangian}
  \delta_{\Lambda} \mathbf{L}  =-d\mathbf{\Xi}[\Lambda]\,,
\end{equation}

\noindent
where $\mathbf{\Xi}[\Lambda]$ is some $(d-1)$-form which depends on $\Lambda$
the equations of motion of the transformed Lagrangian
$\mathbf{L}+\delta_{\Lambda} \mathbf{L}$ will be the same as those derived
from $\mathbf{L}$ and we say that the transformation $\delta_{\Lambda}\varphi$
is a \textit{symmetry} of the theory.

We only consider Lagrangians which are volume forms, such that, when the
fields transform as minus their Lie derivative with respect to a vector field
$\xi$ (\textit{i.e.}~as tensors) $-\mathcal{L}_{\xi}\varphi$, it transforms in
the same way $-\mathcal{L}_{\xi}\mathbf{L} =
d(-\imath_{\xi}\mathbf{L})$. Thus, these transformations are assumed to be
symmetries of the Lagrangian. We refrain from denoting these transformations
with the standard symbol $\delta_{\xi}$ because, as discussed at length in
Refs.~\cite{Elgood:2020svt,Elgood:2020mdx,Elgood:2020nls}, those
transformations must include \textit{induced} or \textit{compensating} gauge
transformation $\delta_{\Lambda_{\xi}}$ whose gauge parameter $\Lambda_{\xi}$
depends on the vector field $\xi$. We will come back to this point later.

If Eq.~(\ref{eq:GaugeTransformationLagrangian}) holds, comparing it with the
equation one obtains by particularizing the first variational formula
Eq.~(\ref{eq:FirstVariationalFormula}) to the $\delta_{\Lambda}$ transformations 

\begin{equation}
  \delta_{\Lambda} \mathbf{L}
  =
  \mathbf{E}_{\varphi}\wedge \delta_{\Lambda}\varphi
  +d\mathbf{\Theta}(\varphi,\delta_{\Lambda}\varphi)\,,
\end{equation}

\noindent
one finds the relation

\begin{equation}
  \label{eq:dJ=ELambda}
  d\left\{\mathbf{\Theta}(\varphi,\delta_{\Lambda}\varphi)
    +\mathbf{\Xi}[\Lambda]\right\}
  =
  -\mathbf{E}_{\varphi}\wedge \delta_{\Lambda}\varphi\,.
\end{equation}

When the infinitesimal parameters of the transformations, $\Lambda$, are
constant, we can write

\begin{equation}
  \delta_{\Lambda}\varphi  = \Lambda^{A}\delta_{A}\varphi\,,
  \hspace{1cm}
  \mathbf{\Xi}[\Lambda] = \Lambda^{A}\mathbf{\Xi}_{A}
\end{equation}

\noindent
where the index $A$ runs over all the independent parameters and
$\delta_{A}\varphi$ and $\mathbf{\Xi}_{A}$ are $\Lambda$-independent. One
obtains the identities

\begin{subequations}
  \begin{align}
  d \mathbf{J}_{A}
   & =
     -\mathbf{E}_{\varphi}\wedge \delta_{A}\varphi\,,
     \\
    & \nonumber \\
    \mathbf{J}_{A}
    & \equiv
      \mathbf{\Theta}(\varphi,\delta_{A}\varphi) +\mathbf{\Xi}_{A}\,,
  \end{align}
\end{subequations}

\noindent
that express the on-shell conservation ($d \mathbf{J}_{A}\doteq 0$\footnote{We
  use $\doteq$ in identities which may only hold on-shell, when
  $\mathbf{E}_{\varphi}=0$.}) of the \textit{Noether currents}
$\mathbf{J}_{A}$. This is the content of Noether's first theorem
\cite{Noether:1918zz}.

When the infinitesimal transformation parameters $\Lambda$ are arbitrary
functions (\textit{i.e.}~in \textit{local} or \textit{gauge} symmetries) the
transformations typically involve their derivatives and the above
factorization is not possible. Eq.~(\ref{eq:dJ=ELambda}), though, implies that
the right-hand side is a total derivative because the left-hand side is, for
any $\varphi$ and any $\Lambda$. Integrating by parts as many times as
necessary one can rewrite it in the form 

\begin{equation}
\label{eq:NoetherSecondTheorem}
\mathbf{E}_{\varphi}\wedge \delta_{\Lambda}\varphi
=
\Lambda^{A}\textbf{N}_{A}[\varphi]+d\mathbf{S}[\Lambda;\mathbf{E}_{\varphi}]\,,
\end{equation}

\noindent
and, since the first term in the right-hand side cannot be rewritten as a
total derivative,  it must identically vanish

\begin{equation}
  \label{eq:Noetheridentities}
  \textbf{N}_{A}[\varphi]
  =
  0\,.
\end{equation}

\noindent
These off-shell identities are known as \textit{gauge} or \textit{Noether
  identities}. They are non-trivial relations between the equations of motion,
implying that they are not all independent and that the theory contains less
degrees of freedom than fields. $\mathbf{S}[\Lambda;\mathbf{E}_{\varphi}]$ is
a $(d-1)$-form proportional to the equations of motion, sometimes called the
\textit{weakly vanishing Noether current} \cite{Barnich:2001jy}.

Using this result, which is the basic content of the second Noether theorem,
in Eq.~(\ref{eq:dJ=ELambda}), we arrive at the off-shell laws establishing the
conservation of the Noether currents $\mathbf{J}[\Lambda]$\footnote{Some
  authors prefer to define the current without the $\mathbf{S}$ factor, making
  the current only closed on-shell. It is, however, impossible to deduce any
  conservation law from this on-shell closed current. We find our definition
  to be more transparent.}

\begin{subequations}
  \begin{align}
    \label{eq:GaugeCurrent}
    \mathbf{J}[\Lambda]
    & \equiv
\mathbf{\Theta}(\varphi,\delta_{\Lambda}\varphi)+\mathbf{S}[\Lambda;\mathbf{E}_{\varphi}]
      +\mathbf{\Xi}[\Lambda]\,,
      \\
    & \nonumber \\
    d \mathbf{J}[\Lambda]
    & = 0\,.
  \end{align}
\end{subequations}

The off-shell closedness is characteristic of the local nature of the
symmetry, and it prevents us from extracting any dynamical conservation
law. However, by virtue of the algebraic Poincar\'e lemma
\cite{Wald:1990mme,Barnich:2000zw},  $\textbf{J}[\Lambda]$ can be
written off-shell as a total derivative

\begin{equation}
  \label{eq:Noetherchargedef}
\textbf{J}[\Lambda]=d\textbf{Q}[\Lambda]\,,  
\end{equation}

\noindent
where $\textbf{Q}[\Lambda]$ is the \textit{gauge} or \textit{Noether}
$(d-2)$-form \textit{charge} associated to the local gauge symmetry.

In general, the Noether charge is not conserved (closed), though. Its
definition Eq.~(\ref{eq:Noetherchargedef}) and the definition of the Noether
current Eq.~(\ref{eq:GaugeCurrent}) imply that

\begin{equation}
  \label{eq:dQLambda}
  d\textbf{Q}[\Lambda]
  =
  \textbf{J}[\Lambda]
  =
  \mathbf{\Theta}(\varphi,\delta_{\Lambda}\varphi)
  +\mathbf{S}[\Lambda;\mathbf{E}_{\varphi}]
  +\mathbf{\Xi}[\Lambda]\,.
\end{equation}

This seems to prevent us from constructing any meaningful conserved charge
associated to gauge symmetries, but there are (at least) two ways out:

\begin{enumerate}
\item Assume the existence of exact symmetries in the solutions whose charges
  you want to define. This approach leads to Page- and Komar-type charges
  which can only be defined in particular solutions.
\item Assume the existence of asymptotic or background symmetries in
  the solutions whose charges you want to define. This is the approach
  of, for instance, Refs.~\cite{Abbott:1981ff,Abbott:1982jh} and leads
  to charges that can be defined for all solutions with a given
  asymptotic behavior and which are defined asymptotically
  (\textit{i.e.}~they are the total charges of the spacetime) and are
  asymptotically conserved.
\end{enumerate}

Here we are going to review the first solution to our problem, which is based
on the observation that the terms in the right-hand side of
Eq.~(\ref{eq:dQLambda}) only vanish under special circumstances:

\begin{enumerate}
\item $\mathbf{S}[\Lambda;\mathbf{E}_{\varphi}]$ vanishes on-shell. Thus, we
  must evaluate it on solutions.
\item Since $\mathbf{\Theta}(\varphi,\delta_{\Lambda}\varphi)$ is always
  linear in $\delta_{\Lambda}\varphi$, it will vanish if the gauge parameters
  generate a transformation that leaves all the fields of the solution
  invariant $\delta_{\Lambda}\varphi=0$. The parameters that have this
  property are called \textit{Killing} or \textit{reducibility parameters}
  $\Lambda_{K}$.
\item For solutions with Killing parameters $\Lambda_{K}$,
  $\delta_{K}\mathbf{L}=0$, which implies that
  $\mathbf{\Xi}[\Lambda_{K}] = d\boldsymbol{w}_{K}$ for some $(d-2)$-form
  $\boldsymbol{w}_{K}$ that, in general, will not be zero..
  
\end{enumerate}

In these conditions, although $d\textbf{Q}[\Lambda_{K}]$ is not zero on-shell,
we can construct a conserved charge combining it with $\boldsymbol{w}_{K}$:

\begin{equation}
\mathbf{K}[\Lambda_{K}] \equiv   -\textbf{Q}[\Lambda_{K}] +\boldsymbol{w}_{K}\,.
\end{equation}

For $(p+1)$-form fields, the on-shell conserved charge
$\mathbf{K}[\Lambda_{K}]$ is related to the so-called \textit{Page
  charges} \cite{Page:1983mke,Marolf:2000cb}.

\subsubsection{Diffeomorphisms}
\label{sec-diffs}

A particular gauge symmetry is the invariance under diffeomorphisms. As we
mentioned before, we assume that all the theories we are considering have it.
We have also warned that the action of the infinitesimal diffeomorphism
generated by the vector field $\xi$ on fields with other gauge freedoms is
minus the Lie derivative supplemented by a compensating gauge transformation
with a $\xi$- and $\varphi$-dependent parameter

\begin{equation}
  \label{eq:deltaxidef}
\delta_{\xi} = -\mathcal{L}_{\xi}+\delta_{\Lambda_{\xi}}\,.
\end{equation}

\noindent
Sometimes the sum of these complementary transformations can be understood as
a \textit{covariant Lie derivative}.\footnote{The first of these covariant Lie
  derivatives were the so-called \textit{spinorial Lie derivatives}
  constructed by by Lichnerowicz and Kosmann in
  Refs.~\cite{kn:Lich,kn:Kos,kn:Kos2} (see also
  Refs.~\cite{Hurley:cf,Figueroa-O'Farrill:1999va}). This construction was
  extended to arbitrary Lorentz tensor-spinor fields in
  Ref.~\cite{Ortin:2002qb} under the name of \textit{Lie-Lorentz derivatives}
  (see also Refs.~\cite{Godina:2003tc,Fatibene2011} for a more mathematically
  rigorous construction). Yang--Mills-type covariant Lie derivatives (this
  includes the case of the Lorentz group and, hence, Lie-Lorentz derivatives)
  where introduced in Ref.~\cite{Ortin:2015hya} for the Abelian case and more
  general and mathematically rigorous constructions have been given in
  Refs.~\cite{Prabhu:2015vua,Bunk:2023ojb}. See also,
  Ref.~\cite{Giotopoulos:2025weu} for a recent review in the context of
  supergravity and Refs.~\cite{Bandos:2023zbs,Bandos:2024pns,Bandos:2024rit}
  for the extension to local supersymmetries.  In the framework of Wald's
  approach to black-hole thermodynamics, Lie-Lorentz derivatives where first
  used in Ref.~\cite{Jacobson:2015uqa} and \cite{Prabhu:2015vua} and these and
  more general covariant Lie derivatives which do not fit in the Yang--Mills
  case where also used in that context in
  \cite{Elgood:2020svt,Elgood:2020mdx,Elgood:2020nls}. See also
  Refs.~\cite{Barbagallo:2026xop,{kn:MO}}). In all cases examined so far, the prescription
  Eq.~(\ref{eq:deltaxidef}) has been shown to be consistent.}

In the simplest cases, if $A$ denotes the gauge field associated to the gauge
symmetry, then

\begin{equation}
  \label{eq:Lambdaxi}
  \Lambda_{\xi}=\imath_{\xi} A-P_{\xi}\,,
\end{equation}

\noindent
where $P_{\xi}$ is the so-called \textit{momentum map}. When $\xi$ generates a
symmetry of the solutions we are considering (in that case we will denote it
by $k$) the condition $\delta_{k}g_{\mu\nu}=0$ implies that $k$ is a Killing
vector of the metric $g_{\mu\nu}$ and $\delta_{k}A=0$ implies that the
momentum map $P_{k}$ must satisfy certain \textit{momentum map equation}
which, for Yang--Mills fields, takes the form

\begin{equation}
\label{eq:YMmomentummapequation}  
\imath_{k}F +\mathcal{D}P_{k}=0\,,  
\end{equation}

\noindent
where $F$ is the Yang--Mills field strength or curvature and $\mathcal{D}$ is
the gauge covariant derivative. Then,

\begin{equation}
  \delta_{\xi}A
  =
  -\mathcal{L}_{\xi}A +\mathcal{D}\Lambda_{\xi}
  =
  -\left(\imath_{\xi}F +\mathcal{D}P_{\xi}\right)
  \stackrel{\xi=k}{=}
  0\,.
\end{equation}

Since the momentum map transforms in the adjoint representation, the induced
gauge parameter $\Lambda_{\xi}$ will transform almost as a gauge field
\cite{Papadimitriou:2005ii}, \textit{i.e.}~under a gauge transformation $g$,
$\Lambda_{\xi}\mapsto g^{-1} \Lambda_{\xi} g+g^{-1} \imath_{\xi} dg$.

It follows from the previous discussions that, under the transformations
Eq.~(\ref{eq:deltaxidef}) the Lagrangian is invariant up to

\begin{equation}
  \delta_{\xi} \mathbf{L}
  =
  -d\left\{ \imath_{\xi} \mathbf{L}+\mathbf{\Xi}[\Lambda_{\xi}] \right\}\,,
\end{equation}

\noindent
where $\mathbf{\Xi}[\Lambda_{\xi}]$ is nothing but the $(d-1)$-form
defined in Eq.~(\ref{eq:GaugeTransformationLagrangian}) evaluated at
the field-dependent gauge parameter $\Lambda_{\xi}$. Following the
procedure outlined above for the general case we arrive at the
following current and charge

\begin{subequations}
\label{eq:DiffCurrent/Charge}
\begin{align}
  \textbf{J}[\xi]
  & =
    \mathbf{\Theta}(\varphi,\delta_{\xi}\varphi)
    +\mathbf{S}[\xi;\mathbf{E}_{\varphi}]
    +\imath_{\xi} \mathbf{L}+\mathbf{\Xi}[\Lambda_{\xi}]\,,
  \\
  & \nonumber \\
  \textbf{J}[\xi]
  & =
    d\textbf{Q}_{NW}[\xi]\,.
\end{align}
\end{subequations}

\noindent
The diffeomorphism charge $\textbf{Q}_{NW}[\xi]$ is known as the
\textit{Noether--Wald (NW) charge} \cite{Wald:1990mme}.

Now, for solutions admitting a Killing vector $k$ we find

\begin{equation}
  d\textbf{Q}_{NW}[k]
  \doteq
  \imath_{k} \mathbf{L}+\mathbf{\Xi}[\Lambda_{k}]\,.
\end{equation}

\noindent
The right-hand side is a total derivative according to the general
argument\footnote{$0=\delta_{k}\mathbf{L} =-d\left\{ \imath_{k}
    \mathbf{L}+\mathbf{\Xi}[\Lambda_{k}] \right\}$. Also, observe that
  the left-hand side is a total derivative. This implies that the
  right-hand side must also be the total derivative of the same object
  up to on-shell closed forms. The construction of on-shell conserved
  forms can be seen as the construction of the cohomology of the
  covariant phase space.}

\begin{equation}
  \label{eq:wkdef}
  \imath_{k} \mathbf{L}+\mathbf{\Xi}[\Lambda_{k}]
  \dot{\equiv}
  d\boldsymbol{w}_{k}\,,
\end{equation}

\noindent
and the on-shell conserved charge that one constructs following the general recipe

\begin{equation}
\mathbf{K}[k] \equiv   -\textbf{Q}_{NW}[k] +\boldsymbol{w}_{k}\,.
\end{equation}

\noindent
is known as the \textit{generalized Komar charge}
\cite{Liberati:2015xcp,Ortin:2021ade,Mitsios:2021zrn,Meessen:2022hcg}.
The calculation of $\boldsymbol{w}_{k}$ can be simplified by the use
of the functional Euler theorem of Ref.~\cite{Cerdeira:2025elp}.
	
\subsection{A well-posed variational problem}
\label{sec-VariationalProblem}

The action of a theory can be defined as a functional of the field
configurations $S[\varphi]$ which is usually constructed as the integral of
the Lagrangian over a spacetime manifold

\begin{equation}
  S[\varphi]
=
  \int_{\mathcal{M}}\mathbf{L}\,.
\end{equation}

Making use of the the first variational formula
Eq.~(\ref{eq:FirstVariationalFormula}) and using Stokes' theorem

\begin{equation}
\label{eq:FirstVariationalFormulaAction}
  \delta S[\varphi]
=
  \int_{\mathcal{M}}\mathbf{E}_{\varphi}\wedge \delta \varphi
  +\int_{\partial\mathcal{M}}\mathbf{\Theta}(\varphi,\delta \varphi)\,.
\end{equation}

If, imposing appropriate boundary conditions on the fields, one can make the
contribution of the boundary term vanish, the action will be extremized by
field configurations that solve the equations of motion
$\mathbf{E}_{\varphi}=0$. When this is possible in a consistent way, we say
that there is a \textit{well-posed variational principle} from which the
classical equations of motion can be derived.\footnote{There is a
  subtle difference between asking the equations of motion to define a well
  posed initial value problem and asking the variational problem to be well
  defined. We refer the reader to \cite{Dyer:2008hb} for more details.}.

As we have discussed, at the classical level, a variational principle is a
completely unnecessary (but esthetically appealing) feature. However, from
the quantum-mechanical point of view

\begin{enumerate}
\item It is imperative to describe theories through actions that admit a
  well-posed variational principle in the above sense.
\item Furthermore, if the value of the action evaluated is physically
  relevant, we want that action to be exactly gauge invariant rather than
  invariant up to total derivatives.\footnote{In the context of holography,
    though, one may want to work with non-gauge-invariant bulk actions
    because they are the duals of CFTs with global anomalies. See, for
    instance, Refs.~\cite{Witten:1998qj,Kraus:2005zm}. } 
\end{enumerate}

Often, the original Lagrangian of the classical field theory does not meet any
of these two requirements and has to be subtly modified while preserving the
classical equations of motion and the symmetries of the theory,
\textit{i.e.}~by adding appropriate total derivatives/boundary terms to
it. These additional terms, however, are relevant in the construction of the
phase space and the Hamiltonian and, obviously, when evaluating the
(Euclidean) action. Our goal in this section will be to study the formulation
of well-defined variational principles satisfying the above requirements
starting with a classical, diff-invariant Lagrangian.
	
Going back to the variational problem
Eq.~\eqref{eq:FirstVariationalFormulaAction}, the variation of the action was
not zero over the solutions of the equations of motion due to the boundary
term $\int_{\partial \mathcal{M}}\mathbf{\Theta}(\varphi,\delta\varphi)$. For
quite some time, the standard solution to this problem was to impose boundary
conditions on the variations of the fields $\delta \varphi$ and to add a
suitable boundary term, a $(d-1)$-form that we will denote by
$\boldsymbol{\ell}$, to cancel exactly this boundary contribution

\begin{equation}
  \mathbf{\Theta}(\varphi,\delta\varphi)
  +\delta\boldsymbol{\ell}\,\stackrel{\partial\mathcal{M}}{=}0\,,  
\end{equation}

\noindent
where $\stackrel{\partial\mathcal{M}}{=} $ denotes equality over the boundary.
However, in \cite{Harlow:2019yfa}, Harlow and Wu argued that this is too
strong of a condition that should be relaxed allowing for a total
derivative\footnote{This precise formula can already be found in
  \cite{Iyer:1995kg}, though.}

\begin{equation}
\label{eq:ConsistencyCondition}
\mathbf{\Theta}(\varphi,\delta\varphi)
+\delta \boldsymbol{\ell}\,
\stackrel{\partial\mathcal{M}}{=}
d\mathbf{D}(\varphi,\delta\varphi)\,.
\end{equation}

This \textit{consistency condition} only needs to hold at the
boundary\footnote{Harlow and Wu argue that it is not natural to impose this
  condition over timelike regions of the boundary, which are interpreted as
  initial/final state surfaces, and that the consistency condition should only
  be applied to the spatial part of the boundary. For our purposes the
  distinction will not be necessary.}, upon the boundary conditions imposed on
$\delta\varphi$ and modulo the equations of motion
$\mathbf{E}_{\varphi}=0$. We cannot resist noting that this equation is
reminiscent of a first variational formula \eqref{eq:FirstVariationalFormula}
for the boundary term.

Sometimes, \cite{Corichi:2013zza,Guo:2025ohn}, an alternative reasoning is
used in order to arrive at Eq.~\eqref{eq:ConsistencyCondition}: the
presymplectic potential can be (non-uniquely) decomposed over the boundary as
$\delta$-exact part, a $d$-exact part and a remainder

\begin{equation}
\label{eq:DecompositionPresympl}
\mathbf{\Theta}(\varphi,\delta \varphi)
=\delta\left( - \boldsymbol{\ell} \right)
+ d\mathbf{D}(\varphi,\delta\varphi)+\textbf{F}(\varphi,\delta \varphi)\,.
\end{equation}

This decomposition is useful because the last term can only be eliminated by
appropriate boundary conditions, and one is left with the consistency
condition \eqref{eq:ConsistencyCondition}.

Boundary conditions typically fall under three classes: Dirichlet, Neumann and
mixed or Robin. Dirichlet boundary conditions fix the dynamical fields at the
boundary, Neumann boundary conditions fix the normal derivative of the
dynamical fields and mixed boundary conditions fixes a combination of both. In
field theory, however, fixing the normal derivative of the field is not
natural and instead one fixes the momenta conjugate to the dynamical field.
In this work we are only going to study Dirichlet-type boundary conditions.

The second requirement (exact gauge invariance of the improved action) reads,
in our notation,

\begin{equation}
\label{eq:GaugeInvarianceCondition}
\delta_{\Lambda} \boldsymbol{\ell}-\mathbf{\Xi}[\Lambda]=d\mathbf{R}[\Lambda]\,,
\end{equation}

\noindent
where $\mathbf{R}[\Lambda]$ is a $(d-2)$-form locally constructed from the
fields of the theory.

A procedure that we will use repeatedly to construct gauge-invariant
boundary terms will be to replace each gauge connection ($A$, say) by
the difference between that gauge connection and a background
connection $\bar{A}$

\begin{equation}
  \vartheta \equiv A-\bar{A}\,.
\end{equation}

This object transforms covariantly (without inhomogeneous terms) under gauge
transformations, since these act in the say way over $A$ and $\bar{A}$.
According to Eq.~(\ref{eq:Lambdaxi}), the parameters of the compensating gauge
transformations in Eq.~(\ref{eq:deltaxidef}) depend on the gauge fields
themselves. However, the parameters of the compensating gauge transformations
acting on $A$ and on $\bar{A}$ must be equal or, at least, must lead to the same
inhomogeneous term in the transformations of these two fields, ensuring that
$\vartheta$ defined above will transform covariantly under induced gauge
transformations. It is not difficult to see that this is possible if

\begin{equation}
  \label{eq:normalization}
  \mathcal{D}(\Lambda_{\xi}-\bar{\Lambda}_{\xi})=0\,,
  \,\,\,\,\,
  \Rightarrow
  \,\,\,\,\,
  \imath_{\xi}(A- \bar{A}) = P_{\xi}-\bar{P}_{\xi}\,.
\end{equation}

\noindent
It is always possible to add a covariantly constant term $C_{\xi}$ in the
left-hand side, but we will adopt the above normalization.

\section{The Hamiltonian}
\label{sec-Hamiltonian}

In this section we will review how to construct the covariant phase
space and a Hamiltonian generating an arbitrary infinitesimal
diffeomorphisms. For a more detailed introduction to the covariant
phase space, see\footnote{For a nice extensive review, see
  Ref.~\cite{Gieres:2021ekc}. See also Ref.~\cite{Golshani:2026lvc}
  for a more advanced recent review.}
Refs.~\cite{Compere:2018aar,Harlow:2019yfa,Julia:2002df}. We will
study diff-invariant theories with gauge symmetries where the
infinitesimal transformations take the generic form
Eq.~(\ref{eq:deltaxidef}), generalizing Harlow and Wu's construction
\cite{Harlow:2019yfa}, making it possible to apply our results to
theories in which the gravitational field is described by the Vielbein,
for instance.  We stress that that we will obtain a well-defined
Hamiltonian for an arbitrary theories of gravity coupled to matter,
without any integrability condition.
		
A Lagrangian, supplemented with a boundary term and boundary conditions such
that the variational problem is well defined, is expected to define a
consistent \textit{prephase space}\footnote{The prefix refers to the fact that
  in the presence of gauge symmetries, the symplectic form will be degenerate,
  and a consistent quotient by the gauge symmetries (symplectic reduction) is
  needed to obtain a symplectic structure on phase space.}.Usually, the
prephase space is defined as the set of initial data required to characterize
a solution of the equations of motion. However, one can define such phase
space in a covariant way, by letting it be the space of all on-shell fields,
\textit{i.e.}~the space of fields solving the equations of motion. Clearly,
these two formulations are equivalent: the evolution of a Cauchy slice of
initial data will \textit{uniquely} determine the whole field history, the
whole solution. Taking all of them, we get the covariant phase space. In the
opposite direction, picking an arbitrary Cauchy slice will select a preferred
set of initial value data: the values of the fields and their (normal)
derivatives on the slice.

This phase space has the structure of a (infinite dimensional) manifold, and
can be equipped with a presymplectic structure: following
Ref.~\cite{Lee:1990nz} we define a \textit{presymplectic current} as the
$(d-1)$-form given by the exterior (vertical) derivative of the presymplectic
potential

\begin{equation}
\label{eq:Presymplecticcurrent1}
\boldsymbol{\omega} (\varphi,\delta\varphi)
\equiv
\delta \mathbf{\Theta}(\varphi,\delta\varphi)\,,
\end{equation}

In components, labeling two independent variations as $\delta_{1,2}$, the
presymplectic current is

\begin{equation}
\label{eq:Presymplecticcurrent2}
\boldsymbol{\omega}(\varphi,\delta_{1}\varphi,\delta_{2}\varphi)
=
\delta_{1} \mathbf{\Theta}(\varphi,\delta_{2}\varphi)
-\delta_{2} \mathbf{\Theta}(\varphi,\delta_{1}\varphi)
-\mathbf{\Theta}(\varphi,[\delta_{1},\delta_{2}]\varphi)\,.
\end{equation}

Then, we define the \textit{presymplectic form} $\Omega_{\Sigma^{d-1}}$ as the integral
of the presymplectic current over a Cauchy hypersurface $\Sigma^{d-1}$

\begin{equation}
\label{eq:PresymplecticForm}
\begin{aligned}
  \Omega_{\Sigma^{d-1}} (\varphi,\delta\varphi)
  & \equiv
    \int_{\Sigma^{d-1}}\boldsymbol{\omega} (\varphi,\delta\varphi)\,,
  \\
  & \\
  \Omega_{\Sigma^{d-1}}(\varphi,\delta_{1}\varphi,\delta_{2}\varphi)
  & =
    \int_{\Sigma^{d-1}}\
    \boldsymbol{\omega}(\varphi,\delta_{1}\varphi,\delta_{2}\varphi)\,.		
\end{aligned}
\end{equation}

Some immediate properties of the presymplectic current and form are:

\begin{itemize}
\item The presymplectic current $\boldsymbol{\omega} (\varphi,\delta\varphi)$
  is $d$-closed on-shell owing to $[\delta,d]=0$ and to the first variational
  formula Eq.~(\ref{eq:FirstVariationalFormula}) and to $\delta^{2}=0$.  Then,
  the presymplectic form $\Omega_{\Sigma^{d-1}} (\varphi,\delta\varphi)$ is
  independent of the choice of Cauchy slice.
\item The presymplectic current $\boldsymbol{\omega} (\varphi,\delta\varphi)$
  is $\delta$-closed and so is the presymplectic form
  $\Omega_{\Sigma^{d-1}} (\varphi,\delta\varphi)$.
\item The presymplectic current $\boldsymbol{\omega} (\varphi,\delta\varphi)$
  is only defined up to $d$-exact terms. This ambiguity is inherited from the
  presymplectic potential $\mathbf{\Theta} (\varphi,\delta\varphi)$.
\item The presymplectic current $\boldsymbol{\omega} (\varphi,\delta\varphi)$
  is only defined up to the addition of $\delta$-exact terms to the
  presymplectic potential $\mathbf{\Theta} (\varphi,\delta\varphi)$.
\end{itemize}

Using the freedom to add $d$- and $\delta$-exact terms to the presymplectic
potential, following Ref.~\cite{Harlow:2019yfa} we define a \textit{dressed
  symplectic current}\footnote{A dressed presymplectic form was also considered
  in \cite{DePaoli:2018erh} in order to match the presymplectic form in tetrad
  variables to the metric variables.}

\begin{equation}
\label{eq:DressedPresymplecticCurrent}
  \boldsymbol{\omega} (\varphi,\delta\varphi)
  \equiv
     \delta\left[
       \mathbf{\Theta}(\varphi,\delta\varphi)
       +\delta \boldsymbol{\ell}
       -d\mathbf{D}(\varphi,\delta\varphi)
  \right]
 =
 \delta\left[  \mathbf{\Theta}(\varphi,\delta\varphi)
   -d\mathbf{D}(\varphi,\delta\varphi) \right]\,,
\end{equation}

\noindent
where we have used $\delta^{2}=0$. The \textit{dressed symplectic form} is its
integral over a Cauchy hypersurface. The main characteristic of these objects is
that they vanish on-shell when restricted to the boundary
$\partial\mathcal{M}$ by virtue of Eq.~(\ref{eq:ConsistencyCondition}). From
now on we will use $\boldsymbol{\omega} (\varphi,\delta\varphi) $ and
$\Omega_{\Sigma^{d-1}} (\varphi,\delta\varphi)$ to denote the dressed
presymplectic current and form.

Evolution due to the diffeomorphism generated by the vector field $\xi$ is
generated in phase space by a \textit{Hamiltonian} $H[\xi]$ implicitly defined
by the property

\begin{equation}
\label{eq:DefinitionHamiltonian}
\Omega_{\Sigma^{d-1}}(\varphi,\delta\varphi,\delta_{\xi} \varphi)
\equiv
-\delta H[\xi]\,.			
\end{equation}
	
In Ref.~\cite{Iyer:1995kg}, it was noticed that the existence of such
a Hamiltonian based on the undressed presymplectic form had some
obstructions related to the $\delta$-integrability of
$\mathbf{\Theta}$, associated to possible radiation flux across the
boundary (see also Ref.~\cite{Compere:2018aar}).  The dressing
proposed by Harlow and Wu overcomes these issues and leads to a
perfectly well-defined Hamiltonian.

Now, we want to rewrite
$\boldsymbol{\omega}(\varphi,\delta\varphi,\delta_{\xi} \varphi)$ as a total
derivative.  Using the definition Eq.~(\ref{eq:DressedPresymplecticCurrent})
and Eq.~(\ref{eq:Presymplecticcurrent2}):

\begin{equation}
\label{eq:HamiltonianDerivation1}
\begin{aligned}
  \boldsymbol{\omega}(\varphi,\delta\varphi,\delta_{\xi} \varphi)
  &  =
    \delta\left[  \mathbf{\Theta}(\varphi,\delta_{\xi}\varphi)
    -d\mathbf{D}(\varphi,\delta_{\xi}\varphi) \right]
  \\
  & \\
  & \hspace{.5cm}
    -\delta_{\xi}\left[\mathbf{\Theta}(\varphi,\delta\varphi)
    -d\mathbf{D}(\varphi,\delta\varphi) \right]
  \\
  & \\
  & \hspace{.5cm}
    -\left[\mathbf{\Theta}(\varphi,[\delta,\delta_{\xi}]\varphi)
    -d\mathbf{D}\left(\varphi,[\delta,\delta_{\xi}]\varphi\right)\right]\,.
\end{aligned}
\end{equation}

The terms involving $\mathbf{D}$ are already a total derivative, so let us
first massage the ``bare'' presymplectic current constructed from
$\mathbf{\Theta}$ using the transformation Eq.~(\ref{eq:deltaxidef})
	
\begin{equation}
\begin{aligned}
\label{eq:BarePresymplecticForm}
  \boldsymbol{\omega}_{\rm bare}(\varphi,\delta\varphi,\delta_{\xi} \varphi)
  & =
    \delta  \mathbf{\Theta}(\varphi,\delta_{\xi}\varphi)
    -\delta_{\xi}  \mathbf{\Theta}(\varphi,\delta\varphi)
    -\mathbf{\Theta}(\varphi,[\delta,\delta_{\xi}]\varphi)
  \\
  & \\
  & =
    \delta  \left\{ \textbf{J}[\xi]-\mathbf{S}[\xi;\mathbf{E}_{\varphi}]-
    \imath_{\xi} \mathbf{L}-\mathbf{\Xi}[\Lambda_{\xi}] \right\}
  \\
  & \\
  & \hspace{.5cm}
    +d\imath_{\xi} \mathbf{\Theta}(\varphi,\delta\varphi)
    +\imath_{\xi} d \mathbf{\Theta}(\varphi,\delta\varphi)
    -\delta_{\Lambda_{\xi}} \mathbf{\Theta}(\varphi,\delta\varphi)
    -\mathbf{\Theta}(\varphi,\delta_{\delta \Lambda_{\xi}}\varphi)
  \\
  & \\
  & =
    d\left\{ \delta \textbf{Q}_{NW}[\xi]
    +\imath_{\xi} \mathbf{\Theta}(\varphi,\delta \varphi) \right\}
    -\delta  \mathbf{S}[\xi;\mathbf{E}_{\varphi}] 
    -\imath_{\xi}\left( \mathbf{E}_{\varphi}\wedge \delta \varphi \right)
  \\
  & \\
  & \hspace{.5cm}
    -\delta_{\Lambda_{\xi}} \mathbf{\Theta}(\varphi,\delta\varphi)
    -\mathbf{\Theta}(\varphi,\delta_{\delta \Lambda_{\xi}}\varphi)
    -\delta \mathbf{\Xi}[\Lambda_{\xi}]\,.
\end{aligned}
\end{equation}
	
In the first step we have used the definition of the NW charge
\eqref{eq:DiffCurrent/Charge}, the definition of the covariant Lie derivative
and the fact that $[\delta,\delta_{\xi}]=\delta_{\delta \Lambda_{\xi}}$,
\textit{i.e.}, a gauge transformation with gauge parameter
$\delta \Lambda_{\xi} $.  The last three terms can be collected into a total
derivative, which we call $-d\textbf{C}[\varphi, \delta \Lambda_{\xi}]$ as
shown in detail in Appendix~\ref{sec-Computation1} and the two remaining terms
which are not total derivatives vanish on-shell.

Thus, on-shell, the bare presymplectic current is a total derivative:

\begin{equation}
\label{eq:BarePresymp}
 \boldsymbol{\omega}_{\rm bare}(\varphi,\delta\varphi,\delta_{\xi} \varphi)
  \doteq
 d\left\{ \delta \textbf{Q}_{NW}[\xi]+\imath_{\xi} \mathbf{\Theta}(\varphi,\delta \varphi)
  -\textbf{C}[\varphi,\delta \Lambda_{\xi}]\right\}\,.
\end{equation}

This result as a direct generalization of the Iyer-Wald construction in which
gauge freedoms have been properly taken care of. This presymplectic form has
been used in \cite{Elgood:2020svt,Ortin:2022uxa} to derive the first law of
black hole mechanics with work terms.

As we can see, the base presymplectic form is not $\delta$-integrable. When we
add the contribution of the dressing terms, we get

\begin{subequations}
\label{eq:DressedPresym}
\begin{align}
  \boldsymbol{\omega}_{\rm bare}(\varphi,\delta\varphi,\delta_{\xi} \varphi)
  & \doteq
    d\mathbf{W}(\varphi,\delta\varphi,\delta_{\xi} \varphi)\,,
  \\
  & \nonumber \\
  \mathbf{W}(\varphi,\delta\varphi,\delta_{\xi} \varphi)
  & =  \delta \textbf{Q}_{NW}[\xi]+\imath_{\xi} \mathbf{\Theta}(\varphi,\delta
    \varphi)-\textbf{C}[\varphi, \delta \Lambda_{\xi}]
\nonumber   \\
  & \nonumber \\
  & \hspace{.5cm}
    +\delta_{\xi} \mathbf{D}(\varphi,\delta\varphi)
    -\delta \mathbf{D}(\varphi,\delta_{\xi}\varphi)
    +\mathbf{D}(\varphi,\delta_{\delta\Lambda_{\xi}}\varphi)\,.
\end{align}
\end{subequations}

On the boundary we can use the consistency condition
Eq.~(\ref{eq:ConsistencyCondition})

\begin{equation}
\label{eq:DressedPresym2}
\begin{aligned}
\mathbf{W}(\varphi,\delta\varphi,\delta_{\xi} \varphi)
  & \stackrel{\partial\mathcal{M}}{\doteq}
    \delta \textbf{Q}_{NW}[\xi]
    -\delta \imath_{\xi} \boldsymbol{\ell}
    +\imath_{\xi} d\mathbf{D}[\varphi,\delta\varphi]
    -\textbf{C}[\varphi, \delta \Lambda_{\xi}]
  \\
  & \\
  & \hspace{.5cm}
    -\imath_{\xi} d \mathbf{D}[\varphi,\delta\varphi]
    +\delta_{\Lambda_{\xi}}\mathbf{D}[\varphi,\delta\varphi]
    -\delta \mathbf{D}[\varphi,\delta_{\xi}\varphi]
    +\mathbf{D}[\varphi,\delta_{\delta\Lambda_{\xi}}\varphi]
  \\
  & \\
  & =
    \delta \left\{
    \textbf{Q}_{NW}[\xi] -\imath_{\xi} \boldsymbol{\ell}
    -\mathbf{D}[\varphi,\delta_{\xi}\varphi]\right\}
  \\
  & \\
  & \hspace{.5cm}
    +\delta_{\Lambda_{\xi}}\mathbf{D}[\varphi,\delta\varphi]
    +\mathbf{D}[\varphi,\delta_{\delta\Lambda_{\xi}}\varphi]
    -\textbf{C}[\varphi, \delta \Lambda_{\xi}]\,.
\end{aligned}
\end{equation}

As shown in Appendix~\ref{sec-Computation2}, the last three terms combine into
$\delta \mathbf{R}[\Lambda_{\xi}]$

\begin{equation}
  \label{eq:otraidentidad}
      \delta_{\Lambda_{\xi}}\mathbf{D}[\varphi,\delta\varphi]
    +\mathbf{D}[\varphi,\delta_{\delta\Lambda_{\xi}}\varphi]
    -\textbf{C}[\varphi, \delta \Lambda_{\xi}]
    \stackrel{\partial\mathcal{M}}{\doteq}
    \delta \mathbf{R}[\Lambda_{\xi}]\,,
\end{equation}

\noindent
and we can $\delta$-integrate the dressed presymplectic current:

\begin{subequations}
  \label{eq:IntegratedPresymplecticForm}
  \begin{align}
    \boldsymbol{\omega}_{\rm bare}(\varphi,\delta\varphi,\delta_{\xi} \varphi)
 &   \stackrel{\partial\mathcal{M}}{\doteq}
   -d\delta \mathbf{H}[\xi]\,,
    \\
 & \nonumber \\
\label{eq:Hamiltonianchargexi}
    \mathbf{H}[\xi]
 & =
   -\textbf{Q}_{NW}[\xi] +\imath_{\xi} \boldsymbol{\ell}
   +\mathbf{D}[\varphi,\delta_{\xi}\varphi]
   -\mathbf{R}[\Lambda_{\xi}]\,,
  \end{align}
\end{subequations}

\noindent
where $\mathbf{H}[\xi]$ is the \textit{Hamiltonian charge} $(d-2)$-form
generalized for fields with gauge freedom\footnote{It is defined up to the
  addition of arbitrary phase space constants, freedom which can be used to
  appropriately normalize it.}.

Now, using Stokes' theorem and taking into account that the boundary of any
Cauchy surface $\partial\Sigma^{d-1}$ must lie inside the boundary of the
whole manifold $\partial\mathcal{M}$ we find the Hamiltonian as an integral
over the boundary of the Cauchy surface:

\begin{equation}
\label{eq:Hamiltonianxi}
  H[\xi]
  =
  \int_{\partial\Sigma^{d-1}}\mathbf{H}[\xi]\,.
\end{equation}

We have obtained a closed-form expression for the Hamiltonian in terms of
conserved charges of the theory and terms directly related to the boundary
conditions imposed.  This is one of the main results of this paper.

Some comments are in order,

\begin{enumerate}

\item Notice hat Harlow and Wu's definition of Hamiltonian
  \cite{Harlow:2019yfa} is exactly recovered when the fields considered do not
  have any gauge freedom.

\item When the diffeomorphism generates a symmetry of all the fields, we
  denote $\xi=k$ and $\delta_{k}\varphi=0$ so that
  $\mathbf{\Theta}(\varphi,\delta_{k}\varphi) = \delta_{k}\boldsymbol{\ell}=0$
  and, hence, $\mathbf{D}(\varphi,\delta_{k}\varphi) = 0$ the Hamiltonian
  charge is somewhat simplified

\begin{equation}
\label{eq:Hamiltonianchargek}
    \mathbf{H}[k]
 =
   -\textbf{Q}_{NW}[k] +\imath_{k} \boldsymbol{\ell} -\mathbf{R}[\Lambda_{k}]\,,  
\end{equation}

\item The Hamiltonian explicitly depends on the boundary conditions chosen
  through $\boldsymbol{\ell}$, $\mathbf{D}[\varphi,\delta \varphi]$ and
  $\mathbf{R}[\Lambda]$.

\item  In Appendix~\ref{sec-Ambiguities} we review the effect of
the ambiguities in the definitions of the charges and boundary terms on this
definition of Hamiltonian charge, and show that it is unaffected by them.

\item The Hamiltonian can be defined as the integral of $\mathbf{H}[\xi]$
  defined by Eq.~(\ref{eq:Hamiltonianxi}) over any $\Sigma^{d-2}$
  surface. However, only when $\Sigma^{d-2}$ lies at the boundary the
  Hamiltonian is the phase space generator of the diffeomorphism, since in the
  derivation we had to use the consistency condition
  Eq.~(\ref{eq:ConsistencyCondition}) to prove
  Eq.~(\ref{eq:DefinitionHamiltonian}). Moreover, the evaluation on any other
  surface one would need to extend $\mathbf{D}$ away from the boundary, where
  it is defined. However, for a Killing vector generating a symmetry of all
  the fields, $\mathbf{H}[k]$ can be defined unambiguously over any
  codimension $2$ surface.

\item The Hamiltonian charge we have defined is not closed, not for Killing
  vectors and not even at the boundary: one can check directly, using the the
  definition of the charges, that\footnote{This fact can also be
  proven directly using the consistency condition
  Eq.~(\ref{eq:ConsistencyCondition}) for $\delta_{\xi}$ transformations:
  \begin{equation}
    \begin{aligned}
\mathbf{\Theta}(\varphi,\delta_{\xi}\varphi)
      +\delta_{\xi} \boldsymbol{\ell}
      -d\mathbf{D}(\varphi,\delta_{\xi}\varphi)
& \stackrel{\partial \mathcal{M}}{=}
  0\,,
      \\
      & \\
\mathbf{\Theta}(\varphi,\delta_{\xi}\varphi)
      -\imath_{\xi}d \boldsymbol{\ell}
      -d\left[ \imath_{\xi}\boldsymbol{\ell}
      +\mathbf{D}(\varphi,\delta_{\xi}\varphi)\right]
      +\delta_{\Lambda_{\xi}}\boldsymbol{\ell}
& \stackrel{\partial \mathcal{M}}{=}
  0\,,
      \\
      & \\
\mathbf{\Theta}(\varphi,\delta_{\xi}\varphi)
      -\imath_{\xi}d \boldsymbol{\ell}
      -d\left\{ \imath_{\xi}\boldsymbol{\ell}
      +\mathbf{D}(\varphi,\delta_{\xi}\varphi)-\mathbf{R}[\Lambda_{\xi}]\right\}
      +\mathbf{\Xi}[\Lambda_{\xi}]
& \stackrel{\partial \mathcal{M}}{\doteq}
  0\,,
      \\
      & \\
      -\imath_{\xi}\left(\mathbf{L}+d \boldsymbol{\ell}\right)
      -d\left\{-\mathbf{Q}_{NW}[\xi] +\imath_{\xi}\boldsymbol{\ell}
      +\mathbf{D}(\varphi,\delta_{\xi}\varphi)-\mathbf{R}[\Lambda_{\xi}]\right\}
& \stackrel{\partial \mathcal{M}}{\doteq}
  0\,,
    \end{aligned}
  \end{equation}
  where we have used Eq.~(\ref{eq:GaugeInvarianceCondition}) in the
  next-to-last step and Eq.~(\ref{eq:DiffCurrent/Charge}) in the last one.
}

\begin{equation}
\label{eq:CllosednessHamiltonian}
d \mathbf{H}[\xi]
\stackrel{\partial\mathcal{M}}{\doteq}
-\imath_{\xi} \left( \mathbf{L} +d\boldsymbol{\ell}\right)\,.
\end{equation}

This equation holds only at the boundary for a general diffeomorphism but it
is true on the bulk of spacetime for Killing vectors.

Just as in the case of the Page and generalized Komar charges, though,
we can infer from the above equation that its right-hand side is a
total derivative on the boundary, because its left-hand side is, and,
if we manage to find a $\mathbf{h}[\xi]\neq \mathbf{H}[\xi]$ such that

\begin{equation}
  \imath_{\xi} \left( \mathbf{L} +d\boldsymbol{\ell}\right)
  \stackrel{\partial\mathcal{M}}{\doteq}
  d\mathbf{h}[\xi]\,, 
\end{equation}

\noindent
we can construct a new conserved $(d-2)$-form charge

\begin{equation}
\tilde{\mathbf{H}}[\xi] \equiv \mathbf{H}[\xi]-\mathbf{h}[\xi]\,,
\end{equation}

\noindent
whose meaning and use will need to be studied.

\item The above expression for the exterior derivative of the Hamiltonian
  charge allows us to prove that it is gauge invariant up to, at most, a total
  derivative that will not contribute when integrated over $(d-2)$-dimensional
  boundary. Performing a gauge transformation of both sides of
  Eq.~(\ref{eq:CllosednessHamiltonian}) we find 

\begin{equation}
\delta_{\Lambda}d \mathbf{H}[\xi]
\stackrel{\partial\mathcal{M}}{\doteq}
-\imath_{\xi} \delta_{\Lambda}\left( \mathbf{L} +d\boldsymbol{\ell}\right)
=
0\,,
\end{equation}

\noindent
by construction, which implies the invariance of  $\mathbf{H}[\xi]$ up to
total derivatives.

In this calculation the consistency condition
Eq.~(\ref{eq:ConsistencyCondition}) is implicitly used and the gauge
transformations of $\mathbf{D}[\varphi,\delta_{\xi}\varphi]$
contribute to the result. This calculation is only justified at the
boundary. However, for Killing vectors
$\mathbf{D}[\varphi,\delta_{k}\varphi]=0$ and the result is valid on
other codimension-2 surfaces.

\end{enumerate}

\subsection{Examples}
\label{sec-Examples}

In order to illustrate our results, in this section we want to compute
the Hamiltonian of several theories with different boundary
conditions. We have to consider diff-invariant theories and, if they
include a metric or Vielbein field, these necessarily have to be
dynamical fields (gravity theories) because, otherwise, the symmetries
of the theory would be restricted to the isometries of a background
metric. Then, we will start by considering the Einstein--Hilbert (EH)
Lagrangian of General Relativity (GR) using the Vielbein as variable
and the construction of an action with the required properties for
Dirichlet boundary conditions.

\subsubsection{General Relativity}
\label{sec-subGrav}

The EH Lagrangian of GR described by Vielbein one-forms $e^{a}$
is,\footnote{We use the conventions of Refs.~\cite{Ortin:2015hya}. In
  particular, we use a mostly minus signature and the Levi--Civita
  spin connection $\omega^{ab}$ and curvature 2-form $R^{ab}$ are
  defined through the structure equations
\begin{subequations}
\begin{align}
  de^{a}
  & =
    \omega^{a}{}_{b}\wedge e^{b}\,,
    \\
      & \nonumber \\
  R^{a}{}_{b}
  & =
    d\omega^{a}{}_{b}-\omega^{a}{}_{c}\wedge \omega^{c}{}_{b}\,. 		
\end{align}
\end{subequations}
}

\begin{equation}
  \label{eq:EHLagrangian}
\mathbf{L} = (-1)^{d-1}\star(e^{a}\wedge e^{b})\wedge R_{ab}\,.
\end{equation}

Under arbitrary perturbations of the Vielbein

\begin{subequations}
\begin{align}
  \delta \mathbf{L}
  & =
    \mathbf{E}_{a}\wedge \delta e^{a}
    +d\mathbf{\Theta}(e,\delta e)\,,
  \\
  & \nonumber \\
  \mathbf{E}_{a}
  & =
    \imath_{a}\star(e^{b}\wedge e^{c})\wedge R_{bc}\,,
  \\
  & \nonumber \\
  \mathbf{\Theta}(e,\delta e)
  & =
    -\star(e^{a}\wedge e^{b})\wedge \delta \omega_{ab}\,.
\end{align}
\end{subequations}

The theory is invariant under local Lorentz transformations

\begin{equation}
  \label{eq:localLorentztransformations}
  \delta_{\sigma}e^{a} = \sigma^{a}{}_{b}e^{b}\,,
  \,\,\,\,\,
  \Rightarrow
  \,\,\,\,\,
  \delta_{\sigma}\omega^{ab}
  =
  \mathcal{D}\sigma^{ab}\,,
  \hspace{1cm}
  \sigma^{ab} = \sigma^{[ab]}\,,
\end{equation}

\noindent
which should be combined with the Lie derivative to construct the covariant
transformations $\delta_{\xi}$ defined in Eq.~(\ref{eq:deltaxidef}) and which
can be regarded as (minus) Lorentz-covariant Lie derivatives or Lie--Lorentz
derivatives \cite{Ortin:2002qb}. The parameters of the compensating Lorentz
transformations are of Yang--Mills type, given by Eq.~(\ref{eq:Lambdaxi}), that
is

\begin{equation}
  \sigma_{\xi}^{ab}
  =
  \imath_{\xi}\omega^{ab}-P_{\xi}{}^{ab}\,,
\end{equation}

\noindent
where, for $\xi=k$, the momentum map satisfies
Eq.~(\ref{eq:YMmomentummapequation}), that is 

\begin{equation}
\imath_{k}R^{ab}+\mathcal{D} P_{k}{}^{ab} =0\,,
\end{equation}

\noindent
which is always solved, at least. by the \textit{Killing bivector}

\begin{equation}
  P_{k}{}^{ab} = \nabla^{a}k^{b} = -P_{k}{}^{ab}\,.
\end{equation}

The explicit form of the $\delta_{\xi}$ transformations for the Vielbein and
spin connection are

\begin{subequations}
  \begin{align}
    \label{eq:deltaxiea}
    \delta_{\xi} e^{a}
    & =
      -\left( \mathcal{D}\xi^{a}+P_{\xi}{}^{a}{}_{b} e^{b}  \right)\,,
    \\
    & \nonumber \\
    \delta_{\xi} \omega^{ab}
    & =
      -\left( \imath_{\xi}R^{ab}+\mathcal{D} P_{\xi}{}^{ab} \right)\,.
  \end{align}
\end{subequations}

The NW charge associated to the invariance under diffeomorphisms
and the Noether charge associated to the invariance under local Lorentz
transformations are

\begin{subequations}
\begin{align}
\label{eq:NWChargeGR}
  \textbf{Q}_{NW}[\xi]
  & =
    (-1)^{d}\star(e^{a}\wedge e^{b})P_{\xi\, ab}
    =
    (-1)^{d}\star d\hat{\xi}\,,
    \\
    & \nonumber \\
\label{eq:LorentzCharge}
  \textbf{Q}[\sigma]
  & =
    (-1)^{d-1}\star(e^{a}\wedge e^{b})\sigma_{ab}\,,
\end{align}
\end{subequations}

\noindent
where $\hat{\xi}=\xi_{\mu}dx^{\mu}$ denotes the 1-form dual to the vector
$\xi=\xi^{\mu}\partial_{\mu}$. The NW charge is conserved on-shell
for Killing vectors because $\boldsymbol{w}_{k}$, defined in
Eq.~(\ref{eq:wkdef}) vanishes identically and coincides with the Komar charge
\cite{Komar:1958wp}.

We proceed to consider how one can formulate a well-posed variational problem
using the EH Lagrangian with Dirichlet boundary conditions.

It is a well-known fact that the boundary term of the EH Lagrangian (the
presymplectic potential) does not vanish when the simplest, Dirichlet-type,
boundary conditions $\delta e^{a} =0$ are imposed because it depends on the
derivatives of $\delta e^{a}$ contained in $\delta \omega^{ab}$. The problem
can be fixed with $\mathbf{D}[\varphi,\delta\varphi]=0$ in an exactly
Lorentz-invariant fashion ($\mathbf{R}[\sigma]=0$) by adding the term

\begin{equation}
    \label{eq:GHYterm}
    \boldsymbol{\ell}_{GHY}
    =
    \star(e^{a}\wedge e^{b})\wedge\vartheta_{ab}\,,
    \hspace{1cm}
    \vartheta^{ab}
    \equiv
    \omega^{ab}-\bar{\omega}^{ab}\,,
\end{equation}

\noindent
where $\bar{\omega}^{ab}$ is some fixed spin connection.

If one chooses $\bar{\omega}^{ab}$ to be the spin connection of the product
metric induced over the boundary, then $\vartheta^{ab}$ becomes the second
fundamental form \cite{Eguchi:1980jx}. Alternatively, if one chooses it to be
the spin connection of a background configuration then
$\boldsymbol{\ell}_{GHY}$ provides an alternative for the
Gibbons--Hawking--York (GHY) term \cite{York:1972sj,Gibbons:1976ue} with
background subtraction \cite{Aros:2005by}.\footnote{For a recent discussion
  with references see Ref.~\cite{Cerdeira:2026uem}. Notice that the GHY
  boundary term is valid for non-null boundaries only. A boundary term for
  null boundaries has been constructed in Ref.~\cite{Parattu:2015gga} (see
  also Ref.~\cite{Chakraborty:2016yna}).}

According to Eq.~(\ref{eq:Hamiltonianchargexi}), using the NW
charge Eq.~(\ref{eq:NWChargeGR}) and the normalization convention
Eq.~(\ref{eq:normalization}), the corresponding Hamiltonian charge is

\begin{equation}
  \begin{aligned}
    \mathbf{H}_{\rm Dirichlet}[\xi]
    & =
      -\textbf{Q}_{NW}[\xi] +\imath_{\xi} \boldsymbol{\ell}_{GHY}
    \\
    & \\
    & =
      -(-1)^{d}\star(e^{a}\wedge e^{b}) \bar{P}_{\xi\, ab}
      +\imath_{\xi}\star(e^{a}\wedge e^{b})\wedge\vartheta_{ab}\,.
  \end{aligned}
\end{equation}

The first term is $\delta$-closed and, since $\mathbf{H}$ is only determined
up to $\delta$-closed terms, it can be ignored. This gives a Hamiltonian
charge that vanishes identically for $\omega^{ab}=\bar{\omega}^{ab}$, which is a
very reasonable normalization condition.

The integral of this charge over the boundary of a Cauchy hypersurface is, in
general, divergent, even in asymptotically flat spacetimes. Thus, proper
renormalization, is needed. If one chooses the $\bar{\omega}^{ab}$ to be the
spin connection of the boundary Vielbein, the integral with the adequate
normalization factor $(16\pi G_{N}^{(d)})^{-1}$ gives precisely the
gravitational charges (ADM mass, angular momentum etc.) when the Killing
vector generates the corresponding transformations (time translations,
rotations etc.) and it is properly normalized on the boundary.  This charge
was also obtained, for instance, in Ref.~\cite{Corichi:2013zza}. A linearized
version was also obtained in Refs.~\cite{Kastor:2004jk,Gomez-Fayren:2023qly}
in the metric and Vielbein formulations, respectively, as a generalized
conserved charge associated to Killing-Yano tensors.

\subsubsection{The Einstein--Maxwell theory}

The Einstein--Maxwell (EM) Lagrangian in $d$ dimensions is the sum of the EH
Lagrangian (\ref{eq:EHLagrangian}) and the Maxwell Lagrangian, \textit{i.e.}

\begin{equation}
  \label{eq:EMLagrangian}
  \mathbf{L}(e,A)
  =
  (-1)^{d-1}\star(e^{a}\wedge e^{b})\wedge R_{ab}
  +(-1)^{d}\tfrac{1}{2}F\wedge\star F\,,
\end{equation}

\noindent
where $F=dA$ is the field strength of the Maxwell field $A=A_{\mu}dx^{\mu}$.

The equations of motion and the presymplectic potential are

\begin{subequations}
\begin{align}
  \delta \mathbf{L}
  & =
\mathbf{E}_{a}\wedge \delta e^{a}
    +\mathbf{E}\wedge\delta A
    +d\mathbf{\Theta}(e,A,\delta e,\delta A)\,,
  \\
  & \nonumber \\
  \mathbf{E}_{a}
  & =
    \imath_{a}\star(e^{b}\wedge e^{c})\wedge R_{bc}
    +\tfrac{1}{2}\left( \imath_{a}F\wedge\star F-F\wedge \imath_{a}\star F \right)\,,
  \\
  & \nonumber \\
  \mathbf{E}
  & =
    -d\star F\,,
  \\
  & \nonumber \\
  \mathbf{\Theta}(e,A,\delta e,\delta A)
  & =
    -\star(e^{a}\wedge e^{b})\wedge \delta \omega_{ab}+
    \star F\wedge \delta A\,.			
\end{align}
\end{subequations}

This theory is invariant under

\begin{enumerate}
\item U$(1)$ gauge transformations under which the Maxwell field
  transforms as a connection

\begin{equation}
  \label{eq:U1transformations}
    \delta_{\Lambda}A = d\Lambda\,. 
\end{equation}

\item Local Lorentz transformations which only act on the vielbein and
  derived objects according to
  Eq.~(\ref{eq:localLorentztransformations}).

\item Diffeomorphisms which act on the vielbein as in
  Eq.~(\ref{eq:deltaxiea}) and on the Maxwell field as

  \begin{equation}
    \delta_{\xi} A
    =
    -\left( \imath_{\xi} F+dP_{\xi} \right)\,,  
  \end{equation}

  \noindent
  once the compensating U$(1)$ gauge transformations with parameter
  $\Lambda_{\xi}$ given in Eq.~(\ref{eq:Lambdaxi}), that is

  \begin{equation}
    \Lambda_{\xi} = \imath_{\xi}A -P_{\xi}\,,  
  \end{equation}

  \noindent
  has been taken into account.
\end{enumerate}

For Killing vectors, $\xi=k$, the momentum map satisfies the momentum
map equation (\ref{eq:YMmomentummapequation}), that is

\begin{equation}
\imath_{k} F+dP_{k}  = 0\,.
\end{equation}

The gauge charge Eq.~(\ref{eq:GaugeCurrent}) associated to the U$(1)$ gauge
symmetry is

\begin{equation}
\label{eq:U(1)charge}
\mathbf{Q}[\Lambda] = (-1)^{d}\Lambda  \star F\,,
\end{equation}

\noindent
and it is conserved on-shell for Killing gauge parameters $d\Lambda_{K}=0$,
\textit{i.e.}~constant parameters. We can always take $\Lambda_{K}=1$.

The NW charge is 

\begin{equation}
\label{eq:MaxwellDiffCharge}
\mathbf{Q}_{NW}[\xi]
 =
(-1)^{d}\star(e^{a}\wedge e^{b})P_{\xi\, ab}
-(-1)^{d}P_{\xi} \star F\,,
\end{equation}

\noindent
and now it is not conserved even on-shell and for Killing vectors.

Following the procedure outlined in Section~\ref{sec-symmetries} we compute
the term $\boldsymbol{w}_{k}$ defined in Eq.~(\ref{eq:wkdef}) and which is
given by





\begin{equation}
  \boldsymbol{w}_{k}
  =
  (-1)^{d-1}\frac{1}{d-2}\left[ P_{k}\star F
      +\tilde{P}_{k} \wedge F\right]\,,
\end{equation}

\noindent
where $\tilde{P}_{k}$ is the magnetic momentum map $(d-4)$-form, defined by
the equation

\begin{equation}
\imath_{k}\star F  +d\tilde{P}_{k}=0\,.
\end{equation}

\noindent
Then, the on-shell conserved generalized Komar charge is given by
\cite{Bardeen:1973gs,Carter:1973rla,Mitsios:2021zrn,Meessen:2022hcg}

\begin{equation}
  \begin{aligned}
    \mathbf{K}[k]
    & =
      -\mathbf{Q}_{NW}[k] + \boldsymbol{w}_{k}
    \\
    & \\
    & =
(-1)^{d}\left\{-\star(e^{a}\wedge e^{b})P_{\xi\, ab}
      +\frac{(d-3)}{(d-2)} P_{\xi} \star F
      +\frac{1}{d-2}\tilde{P}_{k} \wedge F\right\}\,.
  \end{aligned}
\end{equation}

In order to construct an action with a well-posed variational problem we will
profit form the results obtained for vacuum GR in the previous
section and we will only discuss the additional terms that need to be added to
$\boldsymbol{\ell}$s of pure GR to take care of the Maxwell field. Again, we
only consider Dirichlet boundary conditions.

Imposing $\delta A=0$ over the boundary directly makes the part of the
presymplectic potential that depends on the Maxwell field vanish. No
additional terms need to be added to $\boldsymbol{\ell}_{GHY}$ in
Eq.~(\ref{eq:GHYterm}) and $\mathbf{D}=0$.
		
The contribution of the Maxwell field to the Hamiltonian
Eq.~(\ref{eq:Hamiltonianchargexi}) that generates evolution along a vector
$\xi$ with Dirichlet boundary conditions is identical to its contribution to
the NW charge and

\begin{equation}
\label{eq:HamiltonianMaxwellDirichlet}
\mathbf{H}_{Dirichlet}[\xi]
=
\imath_{\xi}\star(e^{a}\wedge e^{b})\wedge\vartheta_{ab}
+(-1)^{d}P_{\xi} \star F\,,
\end{equation}

\noindent
again, up to a $\delta$-closed term. The momentum map $P_{\xi}$ is defined up
to an additive constant and we can use this freedom to make it vanish on the
background solution.

\subsubsection{Minimal 5-dimensional supergravity}
\label{sec-CSTransgression}

We will finish with a simple example of a theory which is only gauge invariant
up to a total derivative: (the bosonic sector of) minimal 5-dimensional
supergravity, which is an extension of the 5-dimensional EM theory with a
Chern--Simons (CS) term. The Lagrangian we are going to consider is
        
\begin{equation}
\label{eq:LagrangianChernSimons}
\mathbf{L} =  \star(e^{a}\wedge e^{b})\wedge R_{ab}
  -\tfrac{1}{2}F\wedge\star F+\gamma F\wedge F\wedge A\,,
\end{equation}

\noindent
where $\gamma$ is an arbitrary constant that takes the precise value
$\gamma=3^{-3/2}$ for the supergravity theory \cite{Cremmer:1980gs}.

This theory is invariant under the local U$(1)$ transformations
Eq.~(\ref{eq:U1transformations}) up to a total derivative

\begin{equation}
  \delta_{\Lambda}\mathbf{L}
  =
  d\left[ \gamma F\wedge F \Lambda \right]
  \equiv
  d\mathbf{\Xi}[\Lambda]\,.
\end{equation}

Under arbitrary variations of the fields

\begin{subequations}
\begin{align}
  \delta \mathbf{L}
  & =
\mathbf{E}_{a}\wedge \delta e^{a}
    +\mathbf{E}\wedge\delta A
    +d\mathbf{\Theta}(e,A,\delta e,\delta A)\,,
  \\
  & \nonumber \\
  \label{eq:EinsteineqsCS}
  \mathbf{E}_{a}
  & =
    \imath_{a}\star(e^{b}\wedge e^{c})\wedge R_{bc}
    +\tfrac{1}{2}\left( \imath_{a}F\wedge\star F-F\wedge \imath_{a}\star F \right)\,,
  \\
  & \nonumber \\
  \label{eq:MaxwelleqsCS}
  \mathbf{E}
  & =
    -d\left[ \star F-3\gamma F\wedge A \right]\,,
  \\
  & \nonumber \\
  \mathbf{\Theta}(e,A,\delta e,\delta A)
  & =
    -\star(e^{a}\wedge e^{b})\wedge \delta \omega_{ab}
   + \left( \star F-2 \gamma F\wedge A \right)\wedge \delta A\,.		
\end{align}
\end{subequations}
	
The charge associated to the $U(1)$ invariance is \cite{Barbagallo:2025qdy}

\begin{equation}
\label{eq:GaugeChargeCS}
\mathbf{Q}[\Lambda] = -\Lambda  \left( \star F-3 \gamma F\wedge A \right)\,,
\end{equation}

\noindent
and it is on-shell conserved for constant $\Lambda$.

The NW charge is

\begin{equation}
\label{eq:DiffChargeCS}
\mathbf{Q}_{NW}[\xi]
 =
-\star(e^{a}\wedge e^{b})P_{\xi\, ab}
+P_{\xi}  \left( \star F-2\gamma F\wedge A \right)\,,
\end{equation}

\noindent
and, yet again, it is not conserved on-shell even for Killing vectors. 
Following the procedure outlined above we can construct the on-shell conserved
generalized Komar charge, given by \cite{Barbagallo:2025qdy}

\begin{equation}
  \label{eq:generalizedKomarCS}
  \mathbf{K}[k]
  =
  \star (e^{a}\wedge e^{b})P_{k\, ab}
  +\tfrac{1}{3} \tilde{P}_{k}\wedge F
  -\tfrac{2}{3}P_{k} \wedge\star F\,,  
\end{equation}

\noindent
where the magnetic momentum 1-form map $\tilde{P}_{k}$ satisfies the equation
\cite{Gibbons:2013tqa}

\begin{equation}
  \label{eq:magneticmomentummapequationCS}
  \imath_{k}\star F
  -6\gamma P_{k}\wedge F +d\tilde{P}_{k}
  \doteq
  0\,.
\end{equation}

Let us now address the problem of constructing an action with a well-posed
variational problem with Dirichlet boundary conditions.

As in the EM case, we do not need to modify the GHY boundary
term $\boldsymbol{\ell}_{GHY}$ in Eq.~(\ref{eq:GHYterm}) in order to have a
well-posed variational problem. However, because of the CS term,
the Lagrangian is gauge invariant up to a total derivative and without
additional boundary terms it is impossible to render the full action gauge
invariant and fulfill the second requirement
Eq.~(\ref{eq:GaugeInvarianceCondition}).  One can get around this problem by
fixing the gauge but we want to make all statements gauge invariant without
breaking the gauge symmetry.  As we are going to see the proper way of dealing
with CS terms is to use a proper background subtraction scheme
through \textit{transgression forms}.

As in the previous cases studied, performing a background subtraction
typically removes the gauge problems, making use of the fact that the
difference of connections transforms tensorially. This is precisely the idea
behind the introduction of \textit{transgression forms}
\cite{Izaurieta:2005vp,Mora:2006ka, Eguchi:1980jx}\footnote{See also
  Ref.~\cite{Cai:2024tyv} where this idea is used in the context of the
  Euclidean path integral.} which act as a consistent background subtraction
scheme for the CS terms that renders the action fully gauge
invariant.

Here we will focus on the 5-dimensional case for concreteness and simplicity,
but the same ideas apply to any dimension. The CS form,
$\mathbf{CS}(A)=F\wedge F\wedge A$ is only locally defined and it satisfies
the formal property\footnote{6-forms vanish identically in 5 dimensions.}

\begin{equation}
  d\mathbf{CS}=F\wedge F\wedge F\,.
\end{equation}

A transgression form $\mathbf{Q}(A,\bar{A})$ depends on an auxiliary,
background connection $\bar{A}$ defined over the same bundle as $A$ and can be
expressed as

\begin{equation}
  \mathbf{Q}(A,\bar{A})
  \equiv
  \vartheta \wedge
  \left( F\wedge F+F\wedge \bar{F}+\bar{F}\wedge \bar{F} \right)\,,
\end{equation}

\noindent
where

\begin{equation}
  \bar{F}\equiv d\bar{A}\,,
  \hspace{1cm}
  \vartheta\equiv A-\bar{A}\,.
\end{equation}

\noindent
$\vartheta$ is now a \textit{globally defined}\footnote{The requirement that
  both $A$ and $\bar{A}$ are defined over the same bundle, \textit{i.e.}~they
  have the same topological structure is necessary for this to be true.}
gauge-invariant 1-form (not a connection). Thus, the transgression form is a
globally well defined 5-form.

Furthermore, the transgression for satisfies the formal property

\begin{equation}
  \label{eq:dQ}
  d\mathbf{Q}(A,\bar{A})=F\wedge F\wedge F-\bar{F}\wedge \bar{F}\wedge \bar{F}\,,
\end{equation}

\noindent
which, on the one hand, expressed the desired notion of background subtraction
and, in the other, guarantees that if we replace by $\mathbf{CS}(A)$ by
$\mathbf{Q}(A,\bar{A})$ the equations of motion will not
change.\footnote{Notice that Eq.~(\ref{eq:dQ}) means that
  \begin{equation}
    \mathbf{Q}(A,\bar{A})
    =
    \mathbf{CS}(A) -\mathbf{CS}(\bar{A})+\text{total derivatives}.
  \end{equation}
} Thus, this discussion leads to replace the original Lagrangian of minimal
5-dimensional supergravity Eq.~(\ref{eq:LagrangianChernSimons}) by the
EM Lagrangian coupled to the above transgression form.  The
Lagrangian we want to consider is, therefore

\begin{equation}
  \label{eq:transgressionLagrangian}
  \mathbf{L}
  \equiv
  \star(e^{a}\wedge e^{b})\wedge R_{ab}
  -\tfrac{1}{2} F\wedge\star F
  +\gamma\theta\left(   F\wedge F+F\wedge \bar{F}+\bar{F}\wedge \bar{F}\right)\,.
\end{equation}

The equations of motion of $e^{a}$ and $A$ are those of the original
theory. However, the presymplectic potential changes and, when we vary
$\bar{A}$ (for instance, when we perform gauge transformations), there is a
coefficient $\mathbf{E}_{\bar{A}}$ to be taken into account:

\begin{subequations}
  \label{eq:transgressionequations1}
\begin{align}
  \mathbf{E}_{\bar{A}}
  & =
    -3\gamma \bar{F}\wedge \bar{F}\,,
  \\
  & \nonumber \\
  \mathbf{\Theta}(\varphi,\delta\varphi)
  & =
    -\star(e^{a}\wedge e^{b})\wedge \delta \omega_{ab}
    \nonumber \\
  & \nonumber \\
  & \hspace{.5cm}
    +\left( \star F-2 F\wedge \gamma\vartheta-\gamma\vartheta\wedge\bar{F} \right)\wedge \delta A
    \nonumber \\
  & \nonumber \\
  & \hspace{.5cm}
    -\left(2\gamma\vartheta \wedge \bar{F}
    +\gamma\vartheta\wedge F \right)\wedge \delta \bar{A}\,.
\end{align}
\end{subequations}

The Lagrangian Eq.~(\ref{eq:transgressionLagrangian}) is exactly invariant
under simultaneous U$(1)$ transformation of $A$ and $\bar{A}$. The
corresponding U$(1)$ gauge charge is
	
\begin{equation}
  \mathbf{Q}[\Lambda]
  =
  -\Lambda\left[ \star F -3\gamma\vartheta \wedge (F+\bar{F}) \right]\,,
\end{equation}

\noindent
but it is not conserved on-shell even for constant gauge parameter $\Lambda$
unless $\bar{F}\wedge \bar{F}=0$. This suggests a solution: adding
(subtracting) to the Lagrangian Eq.~(\ref{eq:transgressionLagrangian}) a
background kinetic term

\begin{equation}
  \label{eq:additionalkineticterm}
  +\tfrac{1}{2} \bar{F}\wedge\bar{\star} \bar{F}\,,
\end{equation}

\noindent
for a fixed metric so that the Einstein equations are not
modified. Eqs.~(\ref{eq:transgressionequations1}) become now

\begin{subequations}
  \label{eq:transgressionequations2}
\begin{align}
  \mathbf{E}_{\bar{A}}
  & =
    \left[\bar{\star}\bar{F}-3\gamma \bar{F}\wedge \bar{F}\right]\,,
  \\
  & \nonumber \\
  \mathbf{\Theta}(\varphi,\delta\varphi)
  & =
    -\star(e^{a}\wedge e^{b})\wedge \delta \omega_{ab}
    \nonumber \\
  & \nonumber \\
  & \hspace{.5cm}
    +\left( \star F-2 F\wedge \gamma\vartheta-\gamma\vartheta\wedge\bar{F} \right)\wedge \delta A
    \nonumber \\
  & \nonumber \\
  & \hspace{.5cm}
    -\left(\bar{\star}\bar{F}+2\gamma\vartheta \wedge \bar{F}
    +\gamma\vartheta\wedge F \right)\wedge \delta \bar{A}\,.
\end{align}
\end{subequations}

The U$(1)$ charge takes now the form

\begin{equation}
  \label{eq:conservedU1chargetransgression}
  \mathbf{Q}[\Lambda]
  =
  -\Lambda\left[ \star F -\bar{\star}\bar{F}
    -3\gamma\vartheta \wedge (F+\bar{F}) \right]\,.
\end{equation}

\noindent
It is manifestly gauge-invariant and conserved on-shell for constant gauge
parameters $\Lambda$ if $\bar{F}$ solves the equations of motion in the
background metric, vanishes for $A=\bar{A}$ and has the original form
Eq.~(\ref{eq:LagrangianChernSimons}) when $\bar{A}=0$. Notice that it cannot
be written as the difference between two conserved charges, even though its
derivative is the difference between two equations of motion.

The additional term Eq.~(\ref{eq:additionalkineticterm}) can be understood as
necessary to normalize the action so that it vanishes for the background
metric and the background vector field, but this requires the subtraction of
a background EH term as well.

Thus, after all this discussion and considerations, we conclude that the
Lagrangian we need is

\begin{equation}
  \label{eq:transgressionLagrangianfinal}
  \mathbf{L}
  \equiv
  \mathbf{L}_{EM}(e,A) -\mathbf{L}_{EM}(\bar{e},\bar{A})
  +\gamma\mathbf{Q}(A,\bar{A})\,,
\end{equation}

\noindent
where $\mathbf{L}_{EM}(e,A)$ is the EM Lagrangian
Eq.~(\ref{eq:EMLagrangian}) in $d=5$ for the vielbein $e^{a}$ and the Maxwell
field $A$, while $\bar{e}^{a}$ and $\bar{A}$ describe a solution of the
original equations of motion. This Lagrangian

\begin{enumerate}
\item Leads to the same equations of motion for the dynamical fields as the
  original one Eq.~(\ref{eq:LagrangianChernSimons}), namely
  Eqs.~(\ref{eq:EinsteineqsCS}) and (\ref{eq:MaxwelleqsCS}).
\item Is gauge-invariant by construction.
\item Vanishes identically when we evaluate it over the background metric and
  Maxwell field $\bar{e}^{a},\bar{A}$.
\item Reduces to the original one Eq.~(\ref{eq:LagrangianChernSimons}) when
  $\bar{A}=0$.
\end{enumerate}

The new features of this Lagrangian are the coefficients of the variations of
the background fields and the presymplectic potential. The former are equal to
the Einstein and Maxwell equations evaluated over the background fields and
vanish identically by assumption. The presymplectic potential is

\begin{equation}
  \begin{aligned}
    \mathbf{\Theta}(\varphi,\delta\varphi)
    & =
      -\star(e^{a}\wedge e^{b})\wedge \delta \omega_{ab}
      +\bar{\star}(\bar{e}^{a}\wedge \bar{e}^{b})\wedge \delta
      \bar{\omega}_{ab}
    \\
    & \\
    & \hspace{.5cm}
    +\left( \star F-2 F\wedge \gamma\vartheta-\gamma\vartheta\wedge\bar{F} \right)\wedge \delta A
    \\
    & \\
    & \hspace{.5cm}
      -\left(\bar{\star}\bar{F}+2\gamma\vartheta \wedge \bar{F}
      +\gamma\vartheta\wedge F \right)\wedge \delta \bar{A}\,.		
  \end{aligned}
\end{equation}

The conserved U$(1)$ charge of this theory is given in
Eq.~(\ref{eq:conservedU1chargetransgression}).

Let us now consider the NW charge. The compensating gauge transformations of
all the fields, dynamical and background, must be the same, which leads to the
relation Eq.~(\ref{eq:normalization}), which we rewrite here for the sake of
convenience for the Maxwell field and the spin connection

\begin{equation}
  \imath_{\xi}(A- \bar{A}) = P_{\xi}-\bar{P}_{\xi}\,,
  \hspace{1cm}
  \imath_{\xi}(\omega^{ab}- \bar{\omega}^{ab}) = P^{ab}_{\xi}-\bar{P}^{ab}_{\xi}\,.
\end{equation}

These relations can be used to replace the background momentum maps by other
quantities. Following the standard procedure we find the NW charge

\begin{equation}
  \begin{aligned}
    \mathbf{Q}_{NW}[\xi]
    & =
    -\star(e^{a}\wedge e^{b})P_{\xi\, ab}
      +\bar{\star}(\bar{e}^{a}\wedge \bar{e}^{b})\bar{P}_{\xi\, ab}
    \\
    & \\
    & \hspace{.5cm}
    +\left[\star F -\gamma\vartheta\wedge \left(2F+\bar{F} \right) \right]P_{\xi} 
    -\left[\bar{\star}\bar{F} +\gamma\vartheta\wedge \left(2\bar{F} +F\right) \right]\bar{P}_{\xi}\,, 
  \end{aligned}
\end{equation}

\noindent
which reduces to the one found in  \cite{Barbagallo:2025qdy} when $\bar{A}=0$,
but is exactly gauge invariant.

Since we only need the GHY boundary term to define a well-posed variational
problem with Dirichlet boundary conditions for the vielbein and Maxwell field,
the Hamiltonian charge is

\begin{equation}
  \label{eq:HamiltonianDTransgression}
  \begin{aligned}
    \mathbf{H}[\xi]
    & =
    -\left[\star(e^{a}\wedge e^{b})-\bar{\star}(\bar{e}^{a}\wedge \bar{e}^{b})
      \right]\bar{P}_{\xi\, ab}
    \\
    & \\
    & \hspace{.5cm}
    +\imath_{\xi}\star(e^{a}\wedge e^{b})\wedge \vartheta_{ab}
    -P_{\xi}\left[ \star F -\bar{\star}\bar{F}
      -3\gamma\vartheta \wedge (F+\bar{F}) \right]
      \\
    & \\
    & \hspace{.5cm}
      -\left[\bar{\star}\bar{F}
      +\gamma\vartheta\wedge \left(2\bar{F} +F\right) \right]
      \imath_{\xi}\vartheta\,.
  \end{aligned}
\end{equation}

This charge vanishes identically when evaluated on the background solution.
The second line of this expression generalizes the Hamiltonian charge of the
EM theory Eq.~(\ref{eq:HamiltonianMaxwellDirichlet}) to account for the
CS term in the action: the second term in that line is the electric
charge of this theory times the momentum map. However, while the terms in the
first line of this expression are $\delta$-exact and can be ignored, not all
those of the third can.

\section{The Euclidean path integral versus the Hamiltonian}
\label{sec-EuclideanPathIntegral}

In Ref.~\cite{Gibbons:1976ue} (see also Ref.~\cite{Hawking:1978jz}) Gibbons
and Hawking proposed to to define the path integral for the theory of quantum
gravity in terms of the Euclidean action $I_{E}[g,\varphi] = -iS[g,\varphi]$
($S[g,\varphi]$ is the Wick-rotated Lorentzian action)

\begin{equation}
\mathcal{Z} =\int   \mathbcal{D}g\mathbcal{D}\varphi\, e^{-I_{E}[g,\varphi]}\,.
\end{equation}

Depending on the quantity one wants to compute, one has to sum over all field
configurations (metric $g$ and matter fields $\varphi$) with given boundary
conditions. In most cases, this is difficult or impossible to do in practice
but a saddle-point approximation in which the value of $\mathcal{Z}$ is
approximated by $e^{-I_{E}[g,\varphi]}$ evaluated over a solution of the
classical equations of motion (a \textit{saddle}). This approximation is
associated to the minimal action principle (the action is minimized by those
solutions) \textit{i.e.}~to an exactly gauge-invariant action of the form

\begin{equation}
\label{eq:EuclideanAction}
I_{E}
=
-i\int_{\mathcal{M}}\big( \mathbf{L}+d\boldsymbol{\ell} \big)\,,		
\end{equation}

\noindent
with a well-posed variational principle, as discussed in
Section~\ref{sec-VariationalProblem}.

When considering stationary black-hole saddles, Gibbons and Hawking observed
that, in order to avoid conical singularities at the position of the
(Lorentzian) horizon, their Euclidean time $\tau$ should be taken to be
periodic and that the value of the period should equal the inverse of the
Hawking temperature

\begin{equation}
  \beta=\frac{1}{T_{H}} = \frac{2\pi}{\kappa}\,,
\end{equation}

\noindent
where $\kappa$ is the surface gravity of the horizon. Then, stationary
black-hole saddles belong to the sate of field configurations with that
periodicity in Euclidean time and $e^{-I_{E}[g,\varphi]}$ can be considered as
a first approximation to a thermal partition function in an statistical
ensemble that depends on other boundary conditions imposed on the fields.

In this section we are interested in the evaluation of the Euclidean action on
stationary black-hole saddles in terms of the Hamiltonian studied in the
previous section.

Ever since the seminal papers Refs.~\cite{Wald:1993nt,Iyer:1995kg} (see also
Refs.~\cite{Garfinkle:2000ms,Cassani:2023vsa,Guo:2025ohn}) there have been
proposals to relate the on-shell Euclidean action to the Hamiltonian, with
different restrictions like absence of matter fields or gauge-fixed
fields. Here we want to generalize the previous results for stationary
black-hole saddles showing, in particular, how some of the problems
encountered when dealing with gauge fields can be overcome using the results
in Section~\ref{sec-Hamiltonian}.

Stationary black-hole spacetimes are characterized by a Killing vector $k$
which generates the horizon and becomes null on it
$k^{2}\stackrel{\mathcal{H}}{=}0$. In the Euclidean solution this means that
$k\stackrel{\mathcal{H}}{=}0$: the only section of the horizon that survives
after the Wick rotation is the bifurcation surface and it is a fixed point of
the Killing vector $k$.

If $\hat{k}=k_{\mu}dx^{\mu}$ denotes the one-form dual to
$k=k^{\mu}\partial_{\mu}$, $\frac{\hat{k}}{k^{2}}$ will exactly be the
one-form dual to the thermal cycle.\footnote{If we are dealing with a
  stationary solution and $\partial_{t}-\Omega\partial_{\phi}$ is the
  horizon-generating Killing vector, then we must identify
  $(t,\phi)\sim (t-i\beta,\phi+i\beta \Omega)$. Alternatively, consider the
  coordinates $\hat{\phi}=\phi+\Omega t$, $\hat{t}=t$, so that
  $(\hat{t},\hat{\phi})\sim (\hat{t}-i\beta,\hat{\phi})$. In these
  coordinates, $\partial_{\hat{t}}=\partial_{t}-\Omega \partial_{\phi}$. Thus
  the $\hat{k}/k^{2}=d\hat{t}+A d\hat{\phi}$ for some function $A$ and when
  integrated over S$^{1}_{\beta}$ gives precisely $-i\beta$.}

Now, for a general volume form $\omega$, and for any non-null vector $k$, the
following identity holds,

\begin{equation}
  \label{eq:omegaidentity}
  \omega=\frac{\hat{k}}{k^{2}}\wedge \imath_{k} \omega\,.
\end{equation}

\noindent
This identity can be used to decompose the integrand of the Lagrangian into a
form orthogonal to the thermal cycle and the 1-form dual to it. However, $k$
goes to zero at the origin, so in order to make the decomposition, we first
need to regulate the integral as follows: let $B_{\epsilon}$ be a region
surrounding the horizon whose size is controlled by the regulator $\epsilon$,
so that $\text{Vol}(B_{\epsilon})\rightarrow 0$ as $\epsilon\rightarrow 0$ and
let $\mathcal{M}_{\epsilon}$ what remains of the full Euclidean manifold
$\mathcal{M}$ when removing $B_{\epsilon}$ from it so

\begin{equation}
\mathcal{M}=B_{\epsilon}\cup \mathcal{M}_{\epsilon}\,.  
\end{equation}

$k\neq 0$ on $\mathcal{M}_{\epsilon}$ and, using the above identity we can
decompose the Euclidean action in two pieces,
	
\begin{equation}
\label{eq:EuclideanActionRegulated}
\begin{aligned}
  I_{E}
  & =
    -i\int_{\mathcal{M}_{\epsilon}}	\big( \mathbf{L}+d\boldsymbol{\ell}
    \big)-i\int_{B_{\epsilon}}	\big( \mathbf{L}+d\boldsymbol{\ell} \big)
  \\
  & \\
  & =
    -i\int_{\mathcal{M}_{\epsilon}}	\frac{\hat{k}}{k^{2}}\wedge\imath_{k}
    \big( \mathbf{L}+d\boldsymbol{\ell} \big)
    -i\int_{B_{\epsilon}}	\big( \mathbf{L}+d\boldsymbol{\ell} \big)\,.
\end{aligned}
\end{equation}

The periodicity of $\tau$ was chosen so as to make the metric regular at the
horizon $k=0$. It follows that the Lagrangian (bulk plus boundary terms)
must me regular at the origin\footnote{Note that this statement does not force
  us to choose a gauge such that the gauge fields become regular at the
  origin. The requirement that the full action is gauge invariant
  Eq.~(\ref{eq:GaugeInvarianceCondition}) already takes care of this issue, so
  if the full action is regular in one gauge, it must be in all gauges.}. This
condition guarantees that in the $\epsilon\rightarrow 0$ limit, the second
term will give a vanishing contribution. Note, however, that now
$\mathcal{M}_{\epsilon}$ has an inner boundary, even though $\mathcal{M}$ does
not.
	
Thus, when $k$ generates translations along the thermal circle, we have the
following decomposition

\begin{equation}
  I_{E}
  =
  -i\int_{\mathcal{M}_{\epsilon}}\frac{\hat{k}}{k^{2}} \wedge \imath_{k}
  \left( \mathbf{L}+d\boldsymbol{\ell} \right)
  =
  -i\int_{S^{1}_{\beta}}\frac{\hat{k}}{k^{2}}
  \int_{\Sigma_{\epsilon}} \imath_{k} \left( \mathbf{L}+d\boldsymbol{\ell} \right)\,,
\end{equation}

\noindent
where $\Sigma_{\epsilon}$ is the hypersurface normal to $k$ in
$\mathcal{M}_{\epsilon}$.

Whenever $\imath_{k} \left( \mathbf{L}+d\boldsymbol{\ell} \right) $ is
independent of the coordinate associated to $k$, we can perform the first
integral along the thermal circle. Note that since in general the objects we
are working with have gauge freedom and we are trying to not fix any
particular gauge, we must be careful. We are working with a classical theory
and we want the action to be meaningful so the boundary terms must ensure that
the full bulk Lagrangian plus boundary term to be exactly gauge invariant (if
not, the values of the action would be completely arbitrary).

Since, by assumption, $\mathbf{L}+d\boldsymbol{\ell}$ is gauge invariant,
$\delta_{k}=-\mathcal{L}_{k}$ and demanding that $k$ annihilates all the
fields automatically implies that $\mathbf{L}+d\boldsymbol{\ell}$ (and thus
$\imath_{k} \left( \mathbf{L}+d\boldsymbol{\ell} \right) $) is independent of
the thermal circle,

\begin{equation}
  I_{E}
  =
  -\beta \int_{\Sigma_{\epsilon}}
  \imath_{k} \left( \mathbf{L}+d\boldsymbol{\ell} \right)\,.	
\end{equation}

It is convenient use the definition of the NW charge
Eq.~\eqref{eq:DiffCurrent/Charge} to rewrite $\imath_{k} \mathbf{L}$ as

\begin{equation}
  I_{E}
  =
  -\beta \int_{\Sigma_{\epsilon}} \left(
    d\mathbf{Q}[k]-\mathbf{\Xi}[\Lambda_{k}]
    -d\imath_{k}\boldsymbol{\ell}+\delta_{\Lambda_{k}}\boldsymbol{\ell} \right),
\end{equation}

\noindent
where we have used the fact that, by assumption,
$\delta_{k}=-\mathcal{L}_{k}+\delta_{\Lambda_{k}}$ annihilates
$\boldsymbol{\ell}$. Using then the second requirement
Eq.~(\ref{eq:GaugeInvarianceCondition}) we obtain the following
solution-independent relation

\begin{equation}
\label{eq:EuclideanActionHamiltonian}
  \begin{aligned}
    I_{E}
    & =
    -\beta \int_{\Sigma_{\epsilon}}\left(
      d\mathbf{Q}[k]-d\imath_{k}\boldsymbol{\ell}+d\mathbf{R}[\Lambda_{k}]\right)
    \\
    & \\
    & =
      \beta \int_{\Sigma_{\epsilon}}d\mathbf{H}[k]
    \\
    & \\
    & =
      \beta \int_{\partial\Sigma_{\epsilon}}\mathbf{H}[k]
    \\
    & \\
    & =
      \beta \int_{S^{(d-2)}_{\infty}} \mathbf{H}[k]
      -\beta\int_{\mathcal{H}} \mathbf{H}[k] \,,
  \end{aligned}
\end{equation}

\noindent
where we have used Eq.~(\ref{eq:Hamiltonianchargek}) to identify the integrand
with the derivative of the Hamiltonian charge and then we have used Stokes'
theorem taking into account that, in general, $\partial \Sigma_{\epsilon}$
will have two boundaries: one at infinity and one at the horizon.

\section{The quantum statistical relation}
\label{sec-QSR}

Recapitulating, the Euclidean path integral computes a thermal partition
function which at leading order is given by the exponential of (minus) the
Euclidean on-shell action with appropriate boundary conditions. Furthermore,
Dirichlet boundary conditions correspond to the grand-canonical ensemble in
which the temperature ($\beta$) and chemical potentials (angular velocities of
the horizon $\Omega$ and electric potentials of the horizon\footnote{The
  actual quantity that plays a role as chemical potential is the
  gauge-invariant difference of potential between the horizon and infinity or
  whatever other relevant reference point is chosen.} $\Phi$) and the path
integral with Dirichlet boundary condition computes the grand
partition function $\mathcal{Z}(\beta,\Omega,\Phi)$.

Introducing the Gibbs free energy

\begin{equation}
  G(\beta,\Omega,\Phi)
  \equiv
  -\frac{1}{\beta}  \log{\mathcal{Z}(\beta,\Omega,\Phi)}\,,
\end{equation}

\noindent
one can show that\footnote{See, for instance, Ref.~\cite{Cassani:2024}.}

\begin{equation}
  G(\beta,\Omega,\Phi)
  =
  E-TS-\Omega J -\Phi Q\,,
\end{equation}

\noindent
so it can be seen as the Legendre transform of the energy $E(S,J,Q)$. Also,
rewriting this expression in the form 

\begin{equation}
  \beta G(\beta,\Omega,\Phi)
  =
  -S- \beta E -(\beta\Omega) J -(\beta \Phi) Q\,,
\end{equation}

\noindent
allows us to see $\beta G$ as the Legendre transform of $-S(E,J,Q)$.

Then, the on-shell Euclidean action with Dirichlet boundary conditions
gives the leading term in the saddle point approximation of

\begin{equation}
  I_{E} \sim -\log{\mathcal{Z}(\beta,\Omega,\Phi)}  \sim
  S+\beta \left(E +\Omega J +\Phi Q\right)\,,
\end{equation}

\noindent
a relation known as the \textit{quantum statistical relation} (QSR).

In Section~\ref{sec-EuclideanPathIntegral} we have shown how to understand the
Euclidean path integral in a geometric way as an integral of conserved charges
constructed in Section~\ref{sec-Hamiltonian}. As a corollary we will see how
the QSR follows naturally from these relations, extending previous results in
a transparent and gauge-invariant way.

The Hamiltonian for Dirichlet boundary conditions evaluated at
infinity will give the linear combination of charges associated to the
Killing vector $k$, which will be the one generating the horizon. It
should be stressed that, since we are using the transformation
Eq.~(\ref{eq:deltaxidef}) that includes compensating or induced gauge
transformations, the gauge charges will also be automatically
included. Indeed, if we consider, for instance, the Hamiltonian charge
of the EM theory Eq.~(\ref{eq:HamiltonianMaxwellDirichlet}) we see
that it has a gravitational term that gives the gravitational charges
and a gauge term consisting of the momentum map and the gauge
charge. The momentum map can be interpreted as the electric potential
$\Phi$ \footnote{When $k=\partial_{t}$. When
  $k=\partial_{t}-\Omega\partial_{\phi}$, it is called the
  \textit{co-rotating} electric potential. For other Killing vectors
  (not relevant in this context) it does not have this
  interpretation.} is constant at infinity or in the horizon and the
integral will give the electric charge times the value of the
potential. The structure of the Hamiltonian of the EM theory is
repeated in more general theories with more gauge fields and more
independent components of the angular momentum.  For simplicity, here
we will just focus on this one.\footnote{We are not considering
  charges associated to scalar fields. these appear in the first law
  of black-hole thermodynamics but not in the Smarr formula
  \cite{Gibbons:1996af,Ballesteros:2023iqb} and their inclusion in our
  results requires further study \cite{kn:CO}.}

The Hamiltonian is the generator of the symmetry associated to $k$ and, almost
by definition, for each Killing vector its integral at infinity will give the
associated charges without any dimension-dependent numerical factors, as it
happens with the Komar charge. Then, for
$k=\partial_{t}-\Omega\partial_{\phi}$ the Hamiltonian charge integrated at
infinity gives

\begin{equation}
\label{eq:HamiltonianInfinity}
\int_{S^{d-2}_{\infty}} \mathbf{H}[k]
=
M-\Omega J+\Phi_{\infty} \mathcal{Q}\,.
\end{equation}

Evaluated at the inner boundary (the horizon), the Hamiltonian gives $TS$,
where $T$ is the temperature and $S$ the entropy and, if the potential has not
been normalized to zero there, a contribution from the electric
charge:\footnote{\label{foot:regulargauge}The Maxwell fields of the
  Reissner--Nordstr\"om or Kerr--Newman black holes are often said to be
  singular on the horizon in certain gauges in which $A^{2}=A_{\mu}A^{\mu}$
  is. When the components of $A$ are regular, the singular behavior is due to
  the singular behavior of the inverse metric.  The reason for studying
  $A^{2}$ instead of the components is that, since $A$ is a 1-form, one has to
  compute a coordinate-independent quantity like $A^{2}$ to see the
  singularity. (Notice that $A$ is not really a 1-form but a connection and
  one should consider gauge- and coordinate-invariant quantities like $F^{2}$
  to assess the singularity of the field. A singularity in $A^{2}$ is just a
  singularity of the expression of the field in a particular gauge.)  Usually,
  a gauge transformation that shifts the time component by a constant and sets
  its value to zero on the horizon is performed to make $A^{2}$ regular
  there. This is the so-called \textit{regular gauge} and, in it $A^{2}$ is
  finite on the horizon. However, the momentum map is a scalar and so is the
  electric potential defined by it. It is regular on the horizon for any
  additive constant and one does not need to compute or use any other
  quantity.}

\begin{equation}
\label{eq:HamiltonianHorizon}
\int_{\mathcal{H}} \mathbf{H}[k]
=
TS+\Phi_{\mathcal{H}} \mathcal{Q}\,.
\end{equation}

Plugging both contributions into \eqref{eq:EuclideanAction}, one directly
obtains the QSR 

\begin{equation}
\label{eq:QuantumStatisticalRelation}
\begin{aligned}
  I_{E}
  & =
    -i\left( \int_{\mathcal{M}} \mathbf{L}
    +\int_{\partial\mathcal{M}}\boldsymbol{\ell} \right)
  \\
  & \\
  & =
    \beta\left( \int_{S^{(d-2)}_{\infty}} \mathbf{H}[k]
    -\int_{\mathcal{H}} \mathbf{H}[k] \right)
  \\
  & \\
  & =
    \beta\left\{\left(M-\Omega J+\Phi_{\infty} \mathcal{Q} \right)
    -\left(TS+\Phi_{\mathcal{H}} \mathcal{Q}  \right)\right\}
  \\
  & \\
  & =
    \beta\left( M-\Omega J-\mu \mathcal{Q} \right)-S
  \\
  & \\
  & =
    -\log \mathcal{Z}(\beta,\Omega,\Phi)\,,
	\end{aligned}
\end{equation}

\noindent
where $\mu=\Phi_{\mathcal{H}}-\Phi_{\infty}$ denotes the chemical potential
associated to the electric charge $\mathcal{Q}$.

\subsection{Minimal 5-dimensional supergravity}
\label{sec-MinimalSugra}

Let us see how this works for minimal 5-dimensional supergravity with the
action Eq.~(\ref{eq:transgressionLagrangianfinal}) constructed with a
transgression form and supplemented with the GHY boundary term
Eq.~(\ref{eq:GHYterm}). The Hamiltonian 3-form charge is given in
Eq.~(\ref{eq:HamiltonianDTransgression}), although we are going to ignore the
$\delta$-exact terms in the first line and we just need to specify the
background fields $\bar{A}$ and $\bar{\omega}^{ab}$.

The background vielbein can be taken as the asymptotic vielbein and
$\bar{\omega}^{ab}$ is just the corresponding spin connection. Notice, though,
that although $e^{a}$ approaches asymptotically $\bar{e}^{a}$,
$\bar{\omega}^{ab}$ does not approach $\omega^{ab}$ and $\vartheta^{ab}$ does
not vanish asymptotically. We take $\bar{A}$ to be the asymptotic limit of $A$
and $\vartheta$ vanishes asymptotically. Then, at infinity, removing the
vanishing $\vartheta$ terms

\begin{equation}
  \beta \int_{S^{3}_{\infty}}\mathbf{H}[k]
  =
  \beta \int_{S^{3}_{\infty}} \left[ \imath_{k}\star(e^{a}\wedge e^{b})\wedge
    \vartheta_{ab}
    +\left(\star F-\star \bar{F}\right) P_{k} \right]\,.		
\end{equation}

In this limit, all the notions of electric charge \cite{Marolf:2000cb}
coincide for this choice of boundary conditions and it is given by the
integral of $\star F-\star \bar{F}$.\footnote{This background subtraction
  would be important in, for instance, asymptotically-G\"odel black holes,
  since the maximally supersymmetric G\"odel solution found in
  Ref.~\cite{Gauntlett:2002nw} has a non-trivial Maxwell field.} Thus,
\begin{equation}
  \beta \int_{S^{3}_{\infty}}\mathbf{H}[k]
  =
  \beta \left( M-\Omega J+\Phi_{\infty} \mathcal{Q} \right)\,.
\end{equation}

At the horizon $k=0$, $\imath_{k} \vartheta$ will vanishes as
well\footnote{Since $\imath_{k} \vartheta$ is gauge invariant, this
  statement is independent of the gauge. Thus there would be no need
  to work in the so-called ``regular gauge'' of the field $A$,
  although, as we have explained in the
  footnote~\ref{foot:regulargauge}, in many cases in which the Maxwell
  field has traditionally been considered to be singular on the
  horizon, it should not have been.} and so does the last term of the
Hamiltonian charge in
Eq.~(\ref{eq:HamiltonianDTransgression}). Furthermore, $P_{k}=\Phi$ is
constant over the horizon\footnote{This is the generalized zeroth law
  of black-hole mechanics, see Ref.~\cite{Elgood:2020svt} and
  references therein.} and the term it multiplies is just the electric
(Page) charge Eq.~(\ref{eq:conservedU1chargetransgression}). Thus,

\begin{equation}
  \beta \int_{\mathcal{H}} \mathbf{H}[k]= S+ \Phi_{\mathcal{H}} \mathcal{Q}\,,
\end{equation}

\noindent
and we recover in a gauge-invariant manner the results of \cite{Cassani:2023vsa}.

Putting both results together we reproduce the QSR
	
\begin{equation}
\label{eq:QSR5dN1Sugra}		
\begin{aligned}
  I_{E} & =
          \beta \int_{S^{(d-2)}_{\infty}} \mathbf{H}[k]
          -\beta\int_{\mathcal{H}} \mathbf{H}[k]
  \\
        & \\
        & =
          \beta\left(M-\Omega J+\Phi_{\infty} \mathcal{Q} \right)
          -\left(S+\beta\Phi_{\mathcal{H}}\mathcal{Q}  \right)
  \\
        & \\
	& =
          \beta\left( M-\Omega J- \mu \mathcal{Q} \right)-S\,,
\end{aligned}
\end{equation}

\noindent
where $\mu =\Phi_{\mathcal{H}}-\Phi_{\infty}$.

\section{Conclusions}
\label{sec-Conclusions}

In this paper we have found a closed expression for the Covariant Phase Space
Hamiltonian. We have shown that one can express the Euclidean action as the
Hamiltonian defined over a $(d-2)$-dimensional boundary and that it is
consistent with the quantum statistical relation in Einstein-Maxwell theories
including Chern--Simons terms.

There are several obvious extensions of these results which we would like to
pursue in future publications: we would like to consider theories with scalar
fields and scalar charges, since these have been shown to occur in the first
law of black-hole mechanics \cite{Gibbons:1996af,Ballesteros:2023iqb}. The
addition of a cosmological constant\footnote{In the holographic context one
  should also be able to relate the holographic renormalization of teh action
  via counterterms with the value of the Euclidean action. See, for instance,
  Ref.~\cite{Awad:2022exw} for an example in minimal, guaged, 5-dimensional
  supergravity.} and, more generally, of terms that include a dimensionful
constant is also a very relevant case that needs special treatment. It has
been shown
\cite{Kastor:2008xb,Kastor:2009wy,Kastor:2010gq,Ortin:2021ade,Meessen:2022hcg,Ballesteros:2024prz,Barbagallo:2026xop}
that these constants play the role of thermodynamical variables in Smarr
formulas and in the first law, generalizing the \textit{black-hole chemistry}
framework \footnote{For reviews with many references see
  Refs.~\cite{Mann:2015luq,Kubiznak:2016qmn}.}.

On the other hand, in many cases, the actual derivation of the Hamiltonian
using the general prescription we have found can be technically difficult:
theories with terms of higher order in the curvature require complicated
boundary terms. Furthermore, one would like to derive the Hamiltonian with
other boundary conditions, related to other statistical ensembles. Finally,
one would like to evaluate the Hamiltonian for relevant classes of solutions
to find, upon use of the quantum statistical relation, explicit expressions
for the entropy and other thermodynamical quantities. Work in this direction
is in progress.

\section*{Disclaimer}

This is a fully AI-free paper. 

\section*{Acknowledgments}
	
JLVC would like to thank Alejandro Ruip\'erez for many interesting and
instructive conversations while at the U.~of Padova, to which he is grateful
for its hospitality. TO would also like to thank him for his comments after
carefully reading the manuscript and for many useful conversations. The work
of JLVC and TO has been supported in part by the MCI, AEI, FEDER (UE) grant
PID2024-155685NB-C21 (``Gravity, Supergravity and Superstrings''
(GRASS)). This publication has been funded within the framework of the R\&D\&I
Project CEX2025-001574-S, funded by MICIU/AEI/10.13039/501100011033. The
research presented in this publication falls within the research line
\textit{Strings and Quantum Gravity}. TO wishes to thank M.M.~Fern\'andez for
her permanent support.


\appendix

\section{The proof of Eq.~(\ref{eq:BarePresymp})}
\label{sec-Computation1}

The proof goes as follows. First we particularize the first variational
formula Eq.~(\ref{eq:FirstVariationalFormula}) to gauge transformations with
gauge parameter $\delta \Lambda_{\xi}$, $\delta_{\delta \Lambda_{\xi}}$

\begin{equation}
\label{eq:Argument}
\begin{aligned}
  \delta_{ \delta{\Lambda_{\xi}}}\mathbf{L}
  & =
    \mathbf{E}_{\varphi}\wedge \delta_{\delta
    \Lambda_{\xi}}\varphi+d\mathbf{\Theta}(\varphi,\delta_{\delta\Lambda_{\xi}}
    \varphi)
  \\
  & \\
  & =
    d\left\{ \mathbf{S}[\delta \Lambda_{\xi};\mathbf{E}_{\varphi}]
    +\mathbf{\Theta}(\varphi,\delta_{\delta\Lambda_{\xi}} \varphi)\right\}\,.		
\end{aligned}
\end{equation}

Next, rewrite $\delta_{\delta \Lambda_{\xi}}$ as
$[\delta,\delta_{\Lambda_{\xi}}]$ and act with it on the Lagrangian

\begin{equation}
\begin{aligned}
  [\delta,\delta_{\Lambda_{\xi}}]\mathbf{L}
  & =
    \delta \left( -d\mathbf{\Xi}[\Lambda_{\xi}] \right)
    -\delta_{\Lambda_{\xi}} \left(\mathbf{E}_{\varphi}\wedge \delta \varphi
  +d\mathbf{\Theta}(\varphi,\delta \varphi) \right)
  \\
  & \\
  & =
    -d\left[ \delta \mathbf{\Xi}[\Lambda_{\xi}]
    +\delta_{\Lambda_{\xi}} \mathbf{\Theta}(\varphi,\delta\varphi) \right]
 -\delta_{\Lambda_{\xi}}\left( \mathbf{E}_{\varphi}\wedge \delta \varphi \right).
\end{aligned}
\end{equation}

Now, the last term vanishes in general, since $\delta \varphi$ transforms
covariantly (even if $\varphi$ is a gauge field), while $\mathbf{E}_{\varphi}$
transforms contravariantly, making the product gauge invariant. Equating both
results we obtain the desired identity:
	
\begin{equation}
\label{eq:Identity}
\delta \mathbf{\Xi}[\Lambda_{\xi}]
+\delta_{\Lambda_{\xi}} \mathbf{\Theta}(\varphi,\delta\varphi)
+\mathbf{\Theta}(\varphi,\delta_{\delta\Lambda_{\xi}} \varphi)
\doteq
\mathbf{C}[\varphi,\delta \Lambda_{\xi} ]\,.
\end{equation}

\section{The proof of Eq.~(\ref{eq:otraidentidad})}
\label{sec-Computation2}

To prove Eq.~(\ref{eq:otraidentidad}) we we need to obtain an alternative
expression for $\mathbf{C}[\varphi,\delta \Lambda_{\xi}]$ over the
boundary. Starting from Eq.~(\ref{eq:Identity}) and using the consistency
condition Eq.~(\ref{eq:ConsistencyCondition}), we find

\begin{equation}
\begin{aligned}
  d\mathbf{C}[\varphi,\delta \Lambda_{\xi}]
  & \doteq
    \delta_{\Lambda_{\xi}}\mathbf{\Theta}(\varphi,\delta \varphi)
    +\mathbf{\Theta}(\varphi,\delta_{\delta \Lambda_{\xi}}\varphi)
    +\delta \mathbf{\Xi}[\Lambda_{\xi}]
  \\
  & \\
  & \stackrel{\partial \mathcal{M}}{=}\,
    -\delta_{\Lambda_{\xi}}\delta \boldsymbol{\ell}
    +d\delta_{\Lambda_{\xi}} \mathbf{D}[\varphi,\delta \varphi]
    -\delta_{\delta \Lambda_{\xi}}\boldsymbol{\ell}
    +d\mathbf{D}[\varphi,\delta_{\delta \Lambda_{\xi}}\varphi]
    +\delta \mathbf{\Xi}[\Lambda_{\xi}]
  \\
  & \\
  & =
    -\delta\left(  \delta_{\Lambda_{\xi}}
    \boldsymbol{\ell}-\mathbf{\Xi}[\Lambda_{\xi}] \right)
    +d\left\{\delta_{\Lambda_{\xi}} \mathbf{D}[\varphi,\delta \varphi]
    +\mathbf{D}[\varphi,\delta_{\delta \Lambda_{\xi}}\varphi]\right\}
  \\
  & \\
  & =
    d\left\{-\delta \mathbf{R}[\Lambda_{\xi}]
    +\delta_{\Lambda_{\xi}} \mathbf{D}[\varphi,\delta \varphi]
    +\mathbf{D}[\varphi,\delta_{\delta \Lambda_{\xi}}\varphi]\right\}\,.
\end{aligned}
\end{equation}
      
We want to emphasize that the use of the consistency condition for the
transformation $\delta_{\delta \Lambda_{\xi}}$ is valid, as even though the
induced gauge transformations $\delta_{\Lambda_{\xi}}$ are not physical
transformations and thus may not satisfy the boundary conditions, since it
arises as the commutator of two transformations which do obey the boundary
conditions,
\textit{i.e.}~$[\delta,\delta_{\xi}]=\delta_{\delta \Lambda_{\xi}}$, it must
also obey the boundary conditions.

We can conclude that on-shell and on the boundary,

\begin{equation}
\label{eq:AlternativeExpressionForC}
\mathbf{C}[\varphi,\delta \Lambda_{\xi}]
\stackrel{\partial\mathcal{M}}{\doteq}
-\delta \mathbf{R}[\Lambda_{\xi}]
+\delta_{\Lambda_{\xi}} \mathbf{D}[\varphi,\delta \varphi]
+\mathbf{D}[\varphi,\delta_{\delta \Lambda_{\xi}}\varphi]\,.
\end{equation}

\section{Ambiguities of the Hamiltonian}
\label{sec-Ambiguities}

Some of the objects that we have defined are defined only up to total
derivatives. In this appendix we study the effect of these ambiguities on the
the Hamiltonian.

The addition of total derivatives and the effect on other quantities that
depend on them can be described by the transformations

\begin{subequations}
\begin{align}
\label{eq:Ambiguities}
  \mathbf{\Theta}(\varphi,\delta\varphi)
  & \rightarrow
    \mathbf{\Theta}(\varphi,\delta\varphi)
    +d\mathbf{X}(\varphi,\delta\varphi)\,,
  \\
  & \nonumber \\
  \mathbf{\Xi}[\Lambda]
  & \rightarrow
 \mathbf{\Xi}[\Lambda]+d\mathbf{Y}[\Lambda]\,,
  \\
  & \nonumber \\
  \boldsymbol{\ell}
  & \rightarrow
    \boldsymbol{\ell}+d\mathbf{Z}\,,
  \\
  & \nonumber \\
  \mathbf{D}(\varphi,\delta\varphi)
  & \rightarrow
    \mathbf{D}(\varphi,\delta\varphi)
    +\delta\mathbf{Z}
    +\mathbf{X}(\varphi,\delta\varphi)
    +d\mathbf{W}(\varphi,\delta\varphi)\,,
  \\
  & \nonumber \\
  \mathbf{R}[\Lambda]
  & \rightarrow
 \mathbf{R}[\Lambda]+\delta_{\Lambda} \mathbf{Z}
    -\mathbf{Y}[\Lambda]+d\mathbf{V}[\Lambda]\,,
  \\
  & \nonumber \\
  \mathbf{Q}_{NW}[\xi]
  & \rightarrow
    \mathbf{Q}_{NW}[\xi]
    +\mathbf{X}(\varphi,\delta_{\xi}\varphi)
    +\mathbf{Y}[\Lambda_{\xi}]
    +d\mathbf{U}[\xi]\,.
	\end{align}
\end{subequations}

It is not difficult to see that the Hamiltonian charge $\mathbf{H}[\xi]$
defined in Eq.~(\ref{eq:Hamiltonianxi}) only
changes by a total derivative, which means that the Hamiltonian, obtained by
integration over a $(d-2)$-dimensional boundary, is exactly invariant.

It is not difficult to see that the Page and generalized Komar charges have the
same property.


\end{document}